\documentclass[aps,pra,twocolumn,amsmath,nofootinbib]{revtex4-1}
\usepackage{epsfig}
\usepackage[colorlinks,linkcolor=blue,anchorcolor=blue,
            citecolor=blue,urlcolor=blue,
            breaklinks=true]{hyperref}
\usepackage{graphics}
\usepackage{color}
\usepackage{graphicx}
\usepackage{dcolumn}
\usepackage{bm}
\usepackage{cases}
\usepackage{appendix}
\usepackage{amsfonts}
\usepackage{epstopdf}
\usepackage{ulem}

\newcommand{\tb}{\textbf}

\begin{document}
\author{Yong-Long Wang$^{1, 2, 3, 5}$}
 \email{wangyonglong@lyu.edu.cn}
\author{Hong-Shi Zong$^{1,4,5}$}
\email{zonghs@nju.edu.cn}
\author{Hui Liu$^{1}$}
\email{liuhui@nju.edu.cn}
\author{Yan-Feng Chen$^{1,3}$}
\email{yfchen@nju.edu.cn}
\address{$^{1}$ Department of Physics, Nanjing University, Nanjing 210093, China}
\address{$^{2}$ School of Physics and Electronic Engineering, Linyi University, Linyi 276000, China}
\address{$^{3}$ National Laboratory of Solid State Microstructures, Department of Materials Science and Engineering, Nanjing University, Nanjing 210093, China}
\address{$^{4}$ Department of Physics, Anhui Normal University, Wuhu, Anhui 241000, China}
\address{$^{5}$ Nanjing Institute of Proton Source Technology, Nanjing 210046, China}

\title{Geometry induced quantum Hall effect and Hall viscosity}
\begin{abstract}
For a particle confined to the two-dimensional helical surface embedded in four-dimensional (4D) Euclidean space, the effective Hamiltonian is deduced in the thin-layer quantization formalism. We find that the gauge structure of the effective dynamics is determined by torsion, which plays the role of U(1) gauge potential, and find that the topological structure of associated states is defined by orbital spin which originates from 4D space. Strikingly, the response to torsion contributes a quantum Hall effect, and the response to the deformation of torsion contributes Hall viscosity that is perfectly presented as a simultaneous occurrence of multiple channels for the quantum Hall effect. This result directly provides a way to probe Hall viscosity.
\bigskip

\end{abstract}
\maketitle

\section{Introduction}
The quantum Hall effect (QHE) was observed in two-dimensional (2D) systems at low temperature and in strong magnetic field~\cite{Klitzing1980New, Tsui1982Two}. For quantum Hall (QH) states, the geometrical and topological features are long-standing projects. A prominent feature is the quantized Hall conductance, which results from the topological characteristics of QH states determined by magnetic singularities~\cite{Thouless1982Quantized}, seen as a transversal response to the electromagnetic field. An important progress of QHE in recent years is Hall viscosity (HV), which originates from the metric perturbation~\cite{Avron1995Viscosity} and is seen as an adiabatic response to a gravitational anomaly~\cite{Can2014Fractional, Wiegmann2018Inner} or a framing anomaly~\cite{Fradkin2015Framing}. Alternatively, the HV is also generated by an inhomogenous electromagnetic field~\cite{Hoyos2012Hall}. Recently, the HV was measured in graphene~\cite{Berdyugin2019Measuring}. For the topological structure of QH sates, the quantized Hall conductances can be realized by parity anomaly~\cite{Haldane1988Model, Wen1991Mean} or by the topological charge of defects~\cite{Brandao2017Inertial}. Without external magnetic field, the quantum anomalous Hall effect~\cite{Nagaosa2010Anomalous, Hasan2010Topological, XueQK2013Experimental, QiXL2016The} shows more geometrical features, and the associated geometric responses can be represented by orbital spin~\cite{Gromov2014Density}. Therefore it becomes interesting to study that a 2D quantum system can display innately the QHE and the HV without external electromagnetic field.

A 2D curved surface embedded in a three-dimensional (3D) Euclidean space $\mathbb{R}^3$ can be characterized by curvature, while the 2D curved surface embedded in a four-dimensional (4D) Euclidean space $\mathbb{R}^4$ should be generally described by curvature and torsion together. The torsion can play the role of gauge potential that was given by Refs.~\cite{Fujii1997Geometrically, Jaffe2003Quantum, Wang2018Geometric, Wang2018E}. As a mathematical connection, the torsion is unphysical and immeasurable~\cite{Chen2008Spin, Chen2009Do}, but it becomes physical and measurable in the presence of orbital spin~\cite{Wang2018Geometric}. The response of orbital spin to torsion leads to many observable universal features. One of the most representative features is that the quantized Hall conductance can be seen as an adiabatic response to torsion. The result can open an access to investigate the QH physics in 4D space~\cite{Lohse2018Exploring,Kraus2018Photonic}.

Since a 4D generalization of QHE was constructed by a mathematically induced gauge field~\cite{Zhang2001A}, the experiments of 4D QHE were proposed in a 2D quasicrystal with a quantized charge pump~\cite{Kraus2013Four}, by arranging the ultracold atoms in a 3D optical lattice~\cite{Price2015Four}, through designing the lattice connectivity with real-valued hopping amplitudes~\cite{Price2020Four}, via considering a non-Hermitian system~\cite{Terrier2020Dissipative}, and so on. In light of the effective gauge field given by mathematical connections~\cite{Zhang2001A}, we try to investigate a 2D curved surface with nonzero torsion by embedding in a 4D space in which torsion plays the role of gauge potential. In the presence of the torsion-induced gauge potential, not only the QHE but also the HV will appear in the considered system. Surprisingly, the HV is directly related to the channels of QHE for the position dependence of torsion.

In the present paper, we will consider a particle confined to a helical surface $\mathbb{S}^2$ embedded in $\mathbb{R}^4$ and deduce the effective Hamiltonian, and further discuss the QHE and the HV that are induced by the geometry intrinsic to $\mathbb{S}^2$. As a practical potential, the effective Hamiltonian can be mapped onto $\mathbb{S}^2$ in $\mathbb{R}^3$ that can provide a way to directly probe the QH physics originating from $\mathbb{R}^4$. The present paper is organized as follows. In Sec. II we deduce the effective Hamiltonian for a particle confined to the helical surface embedded in 4D space and discuss its gauge structure. In Sec. III we discuss the geometric magnetic field and the geometric phase induced by torsion. In Sec. IV we demonstrate the QHE that is seen as an adiabatic response to torsion for charged particles. In Sec. V we find that the HV is seen as a response of orbital spin to the deformation of torsion. Strikingly, the HV is presented as a simultaneous occurrence of multiple channels for quantum Hall conductances. Section VI provides conclusions and discussions.

\begin{figure}[htbp]
  \centering
  \includegraphics[width=0.33\textwidth]{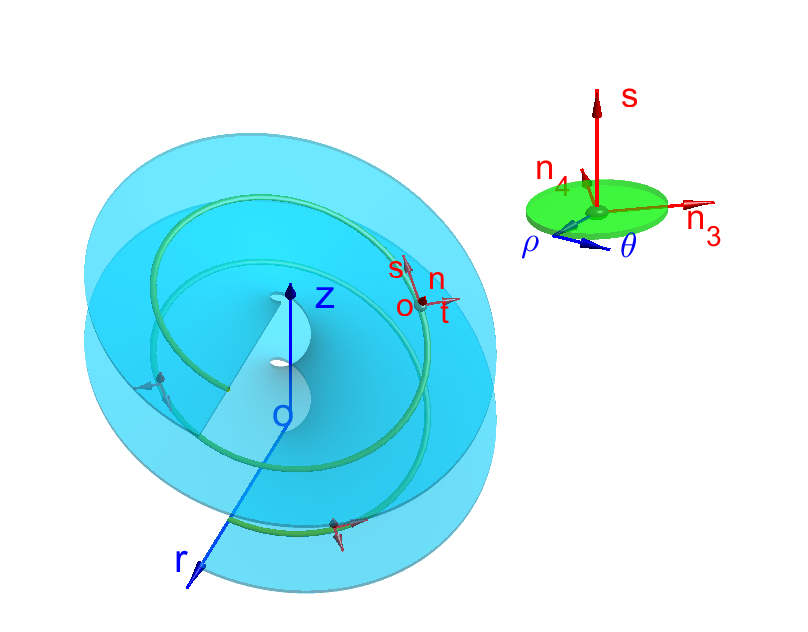}
  \caption{\footnotesize Schematic of a helical surface. $\tb{t}$, $\tb{s}$, and $\tb{n}$ denote three curvilinear coordinate variables. In the upper right figure, $\tb{n}_3$ and $\tb{n}_4$ are both reduced coordinates by introducing a confining potential with the $\rho$ dimension confined.}\label{Fig1}
\end{figure}

\section{Effective Hamiltonian}
A helical surface $\mathbb{S}^2$ embedded in $\mathbb{R}^4$ can be parameterized by
\begin{equation}\label{Surf}
\tb{r}(r,z)=(r, z, r\cos wz,r\sin wz),
\end{equation}
where the $r$ and $z$ coordinates have been mapped onto the helical surface $\mathbb{S}^2$ in a 3D subspace $\mathbb{R}^3$ of $\mathbb{R}^4$. In other words, the relationships of $dt=\sqrt{2}dr$ and $ds=\sqrt{1+w^2r^2}dz$ have been contained in Eq.~\eqref{Surf} with $|\partial\tb{r}/\partial r|=\sqrt{2}$ and $|\partial\tb{r}/\partial z|=\sqrt{1+w^2r^2}$, where $dt$, $dr$, $ds$ and $dz$ are the differential elements of associated coordinate variables, respectively. Here $t$ and $s$ are two tangent coordinate variables of $\mathbb{S}^2$ embedded in $\mathbb{R}^3$ that are sketched in Fig.~\ref{Fig1}, while $r$ and $z$ are two tangent coordinate variables of $\mathbb{S}^2$ embedded in $\mathbb{R}^4$. For describing the helical surface $\mathbb{S}^2$ in $\mathbb{R}^4$, an adapted frame field can be obtained as
\begin{equation}\label{UnitVector}
\begin{split}
& \tb{t}_r=(1, 0, \cos wz, \sin wz),\\
& \tb{t}_z=(0, 1, -wr\sin wz, wr\cos wz),\\
& \tb{n}_3=\frac{1}{\sqrt{2}}(-1, 0, \cos wz, \sin wz),\\
& \tb{n}_4=\frac{1}{\eta}(0, -wr, -\sin wz, \cos wz),
\end{split}
\end{equation}
where $\eta=\sqrt{1+w^2r^2}$, $\tb{t}_r$ and $\tb{t}_z$ are two tangent basis vectors that are used to describe $\mathbb{S}^2$ in $\mathbb{R}^4$, and $\tb{n}_3$ and $\tb{n}_4$ are two normal unit basis vectors that are employed to construct the neighbour space of $\mathbb{S}^2$ in $\mathbb{R}^4$, which are sketched in Fig.~\ref{Fig1}. And thus a point near or on $\mathbb{S}^2$ can be parameterized by
\begin{equation}\label{VSurf}
\tb{R}(r,z, q_3, q_4)=\tb{r}(r,z)+q_3\tb{n}_3(r, z)+q_4\tb{n}_4(r, z).
\end{equation}
Subsequently, the metric tensor $g_{ab}=\partial_a\tb{r}\cdot\partial_b\tb{r}$ $(a,b=r, z)$, the Weingarten curvature tensor $\alpha^k_{ab}=\partial_a\tb{r}\cdot\partial_b\tb{n}^k$, the normal fundamental form $A^{lm}_a=\tb{n}^l\cdot\partial_a\tb{n}^m$ $(k, l, m=3, 4)$ and $G_{ij}=\partial_i\tb{R}\cdot\partial_j\tb{R}$ $(i,j=r,z,3,4)$ can be obtained with Eqs.~\eqref{Surf} and \eqref{VSurf}. The related calculations can be found in Appendix B.

To deduce the effective quantum dynamics for a particle confined to $\mathbb{S}^2$, we first consider a free particle in $\mathbb{R}^4$. The Hamiltonian is
\begin{equation}\label{OHamilton}
{\rm{H}}=-\frac{\hbar^2}{2m^*}\frac{1}{\sqrt{G}}\partial_i\sqrt{G}G^{ij}\partial_j,
\end{equation}
where $G^{ij}$ and $G$ are the inverse and determinant of the metric tensor $G_{ij}$, respectively. According to Eq.~\eqref{OHamilton}, the effective Hamiltonian describing the particle confined to $\mathbb{S}^2$ in $\mathbb{R}^4$ can be obtained by reducing $q_3$ and $q_4$ in the thin-layer quantization formalism~\cite{Jensen1971Quantum, Costa1981Quantum, Jaffe2003Quantum, Wang2016Quantum}. In order to preserve more geometric effects in the effective Hamiltonian, the reduction of $q_3$ and $q_4$ can be accomplished by introducing a confining potential with SO(2) symmetry, $V_c=m^*\omega^2_0\rho^2 /2$, where two polar coordinates $(\rho,\theta)$ are introduced to replace $q_3$ and $q_4$ sketched in the inset of Fig.~\ref{Fig1}. Using the wave function of the ground state of $\rho$~\eqref{A-GroundState} and the formula~\eqref{A-EffHam}, we can obtain the effective Hamiltonian as
\begin{equation}\label{EffHamilton}
{\rm{H}}_{\rm{eff}}=-\frac{\hbar^2}{2m^*}\left[\frac{1}{\eta^2}(\partial_z +i\frac{e_t}{\hbar} \mathcal{A}_z)^2+\frac{1}{2}\partial_r^2\right] +\mathcal{V}_g,
\end{equation}
where $\mathcal{A}_z$ is the geometric gauge potential and $\mathcal{V}_g$ is the geometric potential. Based on the mapping of $r$ and $z$ onto $\mathbb{S}^2$ embedded in $\mathbb{R}^3$, in terms of $t$ and $s$ the effective Hamiltonian $\rm{H}_{eff}$ can be rewritten as
\begin{equation}\label{EffHamilton3}
{\rm{H}}_{\rm{eff}}=-\frac{\hbar^2}{2m^*}\left[(\partial_s +i\frac{e_t}{\hbar} \mathcal{A}_s)^2+\partial_t^2\right] +\mathcal{V}_g,
\end{equation}
where $dt=\sqrt{2}dr$ and $ds=\sqrt{1+w^2r^2}dz$ are considered, and $\mathcal{A}_s=\mathcal{A}_z/\eta$, $e_t=\hbar l$ is a quantized orbital angular momentum, a topological charge (describing the winding number of particle moving around the $\tb{s}$ direction) that results from the embedding of $\mathbb{S}^2$ in $\mathbb{R}^4$. Originally, $e_t$ describes the $\theta$-component motion in the complement space $\mathbb{N}^2$ of $\mathbb{S}^2$ to $\mathbb{R}^4,$ and it is extrinsic to $\mathbb{S}^2$. Converting the extrinsic angular momentum into an intrinsic one, a topological charge is perfectly provided by the SO(2) symmetry of $V_c$~\cite{Wang2018Geometric}. The presence of $e_t$ in Eq.~\eqref{EffHamilton} leads to the wave function owning a certain topological structure. In other words, the intrinsic orbital spin $e_t$ originates from $\mathbb{R}^4$, and the topological property of wave function is also from $\mathbb{R}^4$.

In Eq.~\eqref{EffHamilton} there is a geometric gauge potential $\mathcal{A}_z$ and a geometric scalar potential $\mathcal{V}_g$, which are both important ingredients to confirm the consistency of the effective Hamiltonian $\rm{H}_{eff}$.  The geometric potential~\cite{Costa1981Quantum} $\mathcal{V}_g$ is
\begin{equation}\label{GP}
\mathcal{V}_g=-\frac{\hbar^2}{4m^*}\tau^2,
\end{equation}
where $\tau$ is the torsion of $\mathbb{S}^2$,
\begin{equation}\label{Tau}
\tau=\frac{w}{\sqrt{2(1+w^2r^2)}},
\end{equation}
which is position dependent, a function of $r$. Compared with the given results in~\cite{Dandoloff2004Quantum, Dandoloff2009Geometry}, distinguishably, here $\mathcal{V}_g$ is a locally attractive scalar potential without the repulsive component.

The presence of $\mathcal{A}_z$ in Eq.~\eqref{EffHamilton} determines the U(1) gauge structure of $\rm{H}_{eff}$ due to $\rm SO(2)\simeq U(1)$ in the Abelian case. The strength and representation content of $\mathcal{A}_z$ are defined by a spin connection, which is the normal fundamental form of $\mathbb{S}^2$ with  $\tb{n}_3\cdot\partial_z\tb{n}_4=\tau$, where $\tau$ denotes the twisted angle of the plane spanned by $\tb{n}_3$ and $\tb{n}_4$ around the $\tb{s}$ direction with a unit length increasing along the $z$ axis. Therefore $\mathcal{A}_z$ can be taken as a local rotation of $\mathbb{N}^2$ around a point of $\mathbb{S}^2$ in the $\tb{s}$ direction (sketched in Fig.~\ref{Fig1}). The axis point of rotation does not belong to $\mathbb{N}^2$, so the rotation is a nontrivial singularity, and $\mathbb{N}^2$ has a particular topological structure. Under an infinitesimal rotation in the $\tb{s}$ direction, $\mathcal{R}=e^{-i\theta}$, it is easy to prove that the wave function $|\psi\rangle$ and the geometric gauge potential $\mathcal{A}_z$ satisfy the following transformations
\begin{equation}\label{GTrans}
\begin{split}
&|\psi\rangle\rightarrow |\psi^{\prime}\rangle=\mathcal{R}|\psi\rangle,\\
&\mathcal{A}_z\rightarrow \mathcal{A}_z^{\prime}=\mathcal{A}_z +\partial_z\theta.
\end{split}
\end{equation}
These transformations demonstrate that the geometry intrinsic to $\mathbb{S}^2$ embedded in $\mathbb{R}^4$ can construct the U(1) gauge structure of $\rm{H}_{eff}$. And it is easy to check that $\mathcal{A}_s$ also satisfies the transformation $\mathcal{A}_s\to\mathcal{A}_s^{\prime}=\mathcal{A}_s +\partial_s\theta$. In other words, $\mathcal{A}_s$ can construct the U(1) gauge structure of
the effective Hamiltonian Eq.~\eqref{EffHamilton3} in $\mathbb{R}^3$.

Analogous to the electromagnetic field, the orbital spin $e_t$ plays the role of an electric charge, and $\mathcal{A}_z$ plays the role of an effective electromagnetic field. It is straightforward that the minimal coupling of $\mathcal{A}_z$ is presented in the covariant derivative like that of the electromagnetic field. In the process, the confining potential plays an important role. The SO(2) symmetry of the confining potential is retained in the wave function, which leads the wave function to inherit the topological property of $\mathbb{N}^2$. Particularly, $\mathcal{A}_z$ induced by the geometry intrinsic to $\mathbb{S}^2$ in $\mathbb{R}^4$ is coupled with the orbital spin $e_t$ originating from the $\mathbb{N}^2$ extrinsic to $\mathbb{S}^2$ in $\mathbb{R}^4$. Potentially, the geometry intrinsic to $\mathbb{S}^2$ can be employed to investigate the QH physics in $\mathbb{R}^4$ extrinsic to $\mathbb{S}^2$~\cite{Lohse2018Exploring,Kraus2018Photonic}.

\section{Geometric Magnetic Field and Geometrical Phase}
\begin{figure}[htbp]
  \centering
  \includegraphics[width=0.39\textwidth]{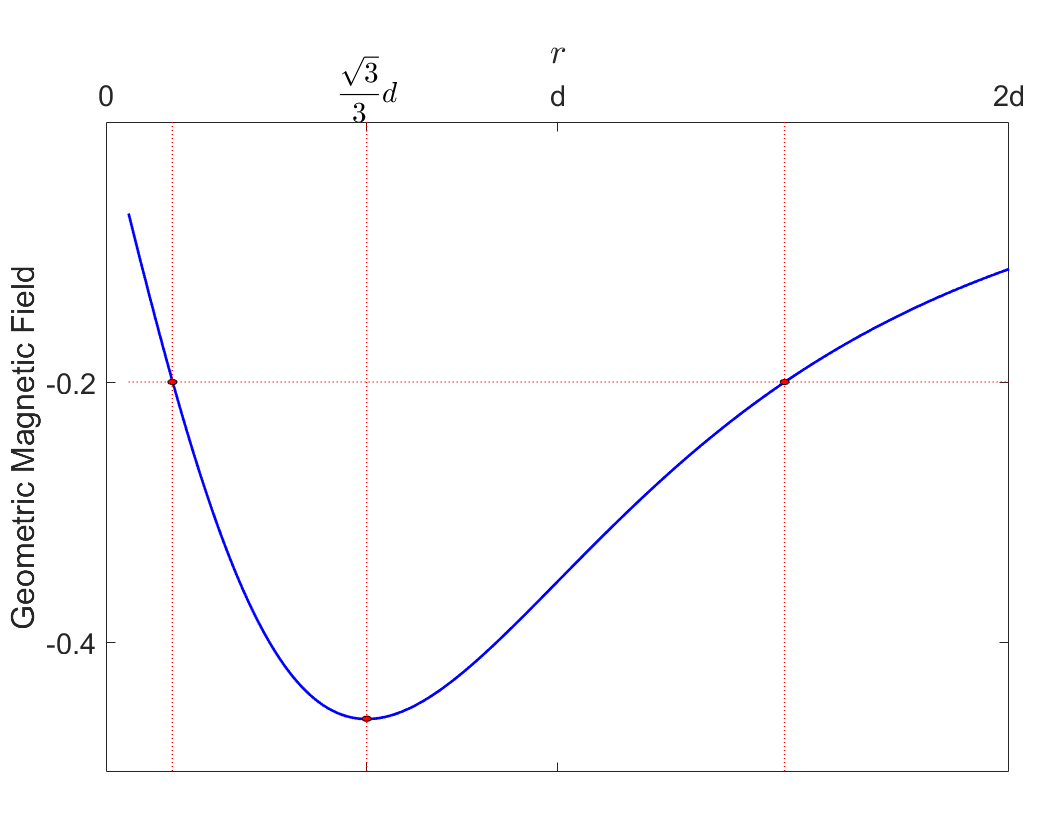}
  \caption{\footnotesize The geometric magnetic field $\frac{\hbar}{e}\mathcal{B}_n$ vs $r$. The scale unit for the $r$ axis is the pitch $d$. The scale unit of the geometric magnetic field is $\frac{\hbar}{e}$ with $d=1$.}\label{Fig2}
\end{figure}

Due to the mapping of $r$ and $z$ onto $\mathbb{S}^2$ in $\mathbb{R}^3$, the geometric gauge potential $\mathcal{A}_z$ in Eq.~\eqref{EffHamilton} in $\mathbb{R}^4$ should be replaced by $\mathcal{A}_s$ in $\mathbb{R}^3$. The subspace $\mathbb{R}^3$ can be spanned by three unit basis vectors $\tb{e}_t$, $\tb{e}_s$, and $\tb{e}_n$. In $\mathbb{R}^3$, the helical surface $\mathbb{S}^2$ can be described by $\tb{r}(r,z)=(z, r\cos wz,r\sin wz)$, the two tangent unit basis vectors $\tb{e}_t$ and $\tb{e}_s$ can be expressed as $\tb{e}_t=(0, \cos wz, \sin wz)$ and $\tb{e}_s=\frac{1}{\eta}(1, -wr\sin wz, wr\cos wz)$, and the normal unit basis vector $\tb{e}_n$ can be obtained as $\tb{e}_n=(wr, \sin wz, -\cos wz)$. With the geometric gauge potential $(0, \mathcal{A}_s, 0)$, the effective magnetic field can be given as
\begin{equation}\label{MagneticField}
\mathcal{B}_n=\partial_t\mathcal{A}_s =-\frac{w^3r}{(1+w^2r^2)^{2}},
\end{equation}
where $dt=\sqrt{2}dr$ is considered, and the subscript $n$ denotes the direction normal to $\mathbb{S}^2$. The geometric magnetic field $\mathcal{B}_n$ is local and nonuniform, since it is a function of $r$, where $r$ is the distance of the point to the $z$ axis. As shown in Fig.~\ref{Fig2}, at $r=\frac{\sqrt{3}}{3}d$ the geometric magnetic field $\mathcal{B}_n$ takes the maximum absolute value, and the particle with certain orbital spin feels the strongest Lorentz-like force. As a result, the particles with different orbital spins are separated. For positive orbital spin, the particles gather to the outer edge of $\mathbb{S}^2$, for negative orbital spin the particles converge to the inner edge. These results can be seen as the response of orbital spin to torsion. The consequence is reminiscent of the QHE. There is a striking feature that the QHE is completely generated by the geometry intrinsic to $\mathbb{S}^2$ in $\mathbb{R}^4$. It is worth mentioning that the geometric magnetic field can have considerable strength when the screw pitch of $\mathbb{S}^2$ takes a suitable value, such as $\frac{\hbar}{e}\mathcal{B}_{n_{max}}\sim -5T$ for $d\sim10nm$. A nonvanishing field strength is a sufficient condition for the geometric gauge potential $\mathcal{A}$ to have a physical effect, but it is not necessary. Even in cases with vanishing field strength, global Aharonov-Bohm effects can exist when the constraint hypersurface has nonvanishing torsion~\cite{Takagi1992}.

In the presence of $\mathcal{A}_s$, the particle with orbital spin confined to $\mathbb{S}^2$ moving along the $s$-direction will gain an additional geometrical phase as
\begin{equation}
\Delta\phi_g=\int_{s_0}^{s_e}l\mathcal{A}_sds=\int_{z_0}^{z_e} l\tau dz,
\end{equation}
where $s_0$ and $s_e$ are the start value and the end value of the integral variable $s$, and $z_0$ and $z_e$ are the start value and the end value of the integral variable $z$, respectively. Obviously, the geometrical phase has $\Delta\phi_g=l\tau$ for a unit length of the $z$ axis. Its sign depends on the sign of the orbital spin, and its value increases by increasing $w$, while it decreases when $r$ takes a larger value. Therefore, the geometrical phase can be adjusted by designing the geometry and size of $\mathbb{S}^2$.

\section{Quantum Hall Effect}
In the geometric magnetic field $\mathcal{B}_n$, the particle with orbital spin feels a Lorentz-like force, $\mathcal{F}=\tb{e}_t ev_sl\mathcal{B}_n$, where $v_s$ is the $s$ component of velocity. The force direction is $\tb{e}_t$ for a positive orbital spin, $-\tb{e}_t$ for a negative one, and the force strength is proportional to the orbital spin. Therefore the particles with positive orbital spin accumulate to the outer side edge of $\mathbb{S}^2$, and the particles with negative orbital spin move to the inner side edge. The redistribution of charged particles will generate an electric field $\mathcal{E}_t$ that is inhomogeneous. In the inhomogeneous electric field $\mathcal{E}_t$, a charged particle feels Coulomb force. The force is position dependent. Finally, there is a force balance $e\mathcal{E}_r=ev_s l\mathcal{B}_n$. The $s$ component current can be expressed as $j_s=e nv_s$ and can be described by $j_s=\sigma_{sr}\mathcal{E}_r$. The Hall conductivity~\cite{Hoyos2012Hall} can be deduced as
\begin{equation}\label{HallCond}
\sigma_{sr}=\frac{e}{\Phi_0}\nu,
\end{equation}
where $\Phi_0=h/e$ is the effective flux quantum, which is the effective magnetic flux contained within the area $2\pi\ell_B^2$, $\nu=n 2\pi \ell_B^2$ denotes a filling factor describing the number of particles coupled to flux quantum $\Phi_0$, wherein $n$ is the particle density, and $\ell_B=\sqrt{1/l\mathcal{B}_n}$ stands for an effective magnetic length.

It is worthwhile to notice that the present QHE is purely induced by torsion. The topological properties of QH states are provided by the presence of orbital spin in $\rm{H}_{eff}$. Those nontrivial properties are determined by the ground states of $\rho$ that are nontrivial irreducible fundamental representations of SO(2). In the presence of the geometric gauge potential, the Landau-like energy level splitting is exhibited and the energy levels increase as the torsion increases. In the case of vanishing torsion, the states below the Fermi energy $E_F$ are fully occupied at zero temperature. By adiabatically increasing torsion, the energy level for the maximum orbital spin will first reach $E_F$, and a channel for the QH current appears. When the torsion continues to increase, the energy for the next orbital spin will reach $E_F$ again, and a new channel will appear again.

Specifically, the orbital spin provides channels for particles moving along the $s$ direction without resistance, and the torsion of $\mathbb{S}^2$ plays the role of a key to open the channels. The orbital spin transport, driven by an adiabatic change of the torsion, is a fundamental probe of QH states with topological characterization, complementary to the more familiar electromagnetic response. The orbital spin can highlight the topological properties of the QH states and can encode the external geometric characterization in the intrinsic degree of freedom, which can serves as an ideal setting to probe their geometric properties. In other words, the intrinsic orbital spin can be a useful tool to probe the geometric responses and to find more subtle features of QH states.

\section{Hall Viscosity}
 The HV is determined by the nondissipative part of the stress response to metric perturbation~\cite{Avron1995Viscosity}, and it can be also created by an inhomogeneous electric field~\cite{Hoyos2012Hall} or by special boundary conditions~\cite{Gromov2016Boundary}. The metric perturbation leads to the quantum geometry of the guiding center~\cite{Haldane2009Hall} that is related to the HV. In the present paper, the torsion of $\mathbb{S}^2$ has the normal fundamental form in Eq.~\eqref{B-NormConnect} that is antisymmetric in the metric tensor $G_{ij}$ and that describes the nondissipative part of the stress response. As an Abelian SO(2) orbit connection, the geometric gauge potential $\mathcal{A}_s$ plays the role of the gravitational Abelian Chern-Simons action~\cite{Gromov2014Density} to generate the HV.

Specifically, for the charged particle with orbital spin confined to $\mathbb{S}^2$, the geometric gauge potential generates an effective magnetic field, and there are torsion-induced Landau levels that are macroscopically degenerate multiplets. In the presence of the geometric gauge potential, the dynamical momentum of the particle confined to $\mathbb{S}^2$ can be expressed as $\pi_a\equiv p_a+\hbar l\mathcal{A}_a$, with $[x^a, p_b]=i\hbar\delta^a_b$ and $[x^a, x^b]=[p_a, p_b]=0$~\cite{Haldane2009Hall}, where $x^a (a=t, s)$ describes two local coordinates of $\mathbb{S}^2$. The guiding centers are $X^a=x^a-(\ell^2_B/\hbar)\varepsilon^{ab}\pi_b$, and they commute with the dynamical momenta. As a consequence, we can deduce the noncommutative relationship for the guiding centers in the following form
\begin{equation}
[X^a, X^b]=i\varepsilon^{ab}\ell_B^2.
\end{equation}
In the above calculations, the geometric magnetic fields $\mathcal{B}_n=\partial_t\mathcal{A}_s$ and $\ell_B=\sqrt{1/l\mathcal{B}_n}$ are considered. The noncommutativity of the guiding centers can be also deduced by the geometry-induced noncommutativity of two different components of momentum~\cite{Liu2017Curvature}. In light of the charge conservation law, we can obtain
\begin{equation}\label{HallViscosity}
\eta^A=\frac{l}{2\pi\ell_B^2},
\end{equation}
which describes the HV. This result is in full agreement with that given by Read~\cite{Read2009NonAbelian} as $\bar{n}=\frac{1}{\pi\ell_B^2}$, which denotes the charged particle density.

\begin{figure}[htbp]
  \centering
  \includegraphics[width=0.39\textwidth]{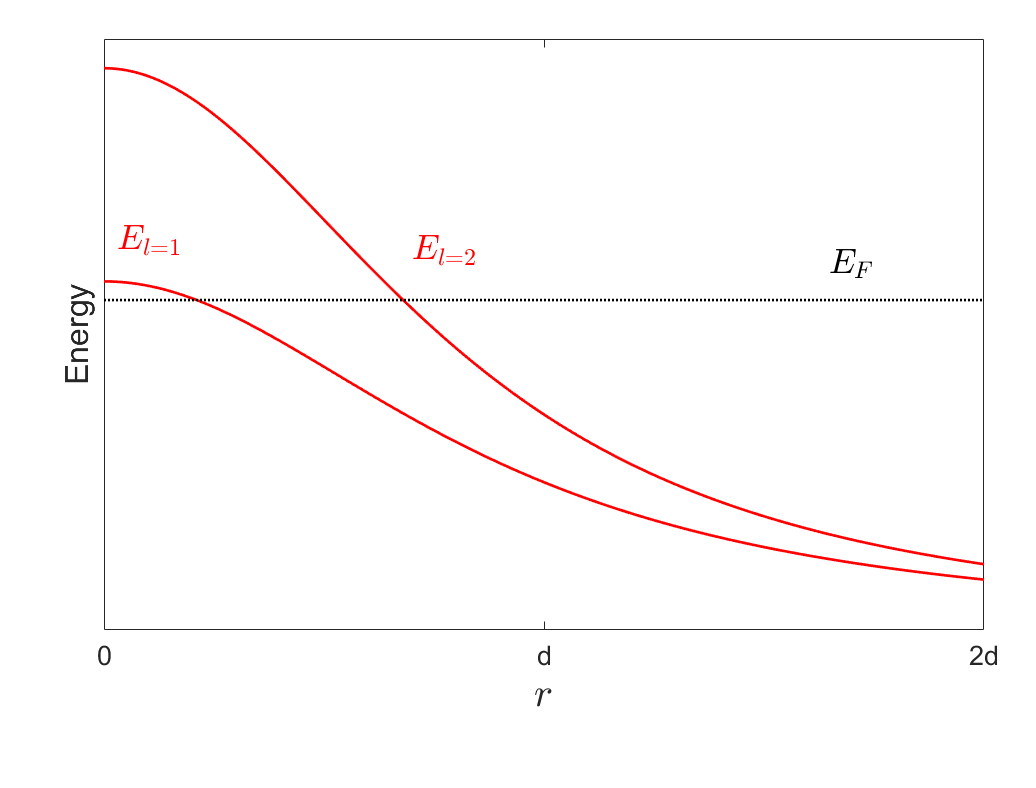}
  \caption{\footnotesize With a certain twisted coefficient, two Landau-like energies cross with $E_F$ at different positions of $r$.}\label{Fig3}
\end{figure}

In what follows, we reconsider the effective Hamiltonian Eq.~\eqref{EffHamilton3}, and write the associated Schr\"{o}dinger equation as
\begin{equation}\label{SchEq}
-\frac{\hbar^2}{2m^*}\left(\partial_s+i\frac{e_t}{\hbar}\mathcal{A}_s\right)^2\psi -\frac{\hbar^2}{2m^*}\partial_t^2\psi-\frac{\hbar^2}{4m^*}\tau^2\psi =E\psi,
\end{equation}
where the term $i\frac{e_t}{\hbar}\mathcal{A}_s$ plays the role of the minimal coupling of a U(1)-like gauge field. Using the ansatz $\psi(r,z)=f(r)\phi(z)$~\cite{Dandoloff2009Geometry}, the $z$ component of the effective Schr\"odinger equation can be given by
\begin{equation}\label{SEz}
-\frac{\hbar^2}{2m^*}\frac{1}{\eta^2}\left(\frac{d}{dz}+i\frac{e_t}{\hbar}\mathcal{A}_z\right)^2\phi(z)=E_0\phi(z).
\end{equation}
With a solution $\phi(z)=e^{ik_zz}$ of Eq.~\eqref{SEz} without $i\frac{e_t}{\hbar}\mathcal{A}_s$, $E_0$ in Eq.~\eqref{SEz} can be expressed as
\begin{equation}\label{Ez}
E_0=\frac{\hbar^2}{2m^*}\frac{1}{\eta^2}(k_z+l\tau)^2,
\end{equation}
where $k_z$ is the $z$ component of momentum and $e_t=\hbar l$ and $\mathcal{A}_s=\tau/\eta$ have been considered. There is a periodicity in the $z$ axis for the present system, and the azimuthal angle twisted around $z$ is described by $w z$. Thus the $z$ component of angular momentum can be expressed as $L_z=-i\frac{\hbar}{w}\partial_z$ with eigenvalue $\hbar m$, and the momentum $k_z$ is quantized as $k_z=mw$, $m\in N$. Here $\mathcal{A}_s$ plays the role of gauge potential which minimally couples with the topological charge $\hbar l$ to generate the Landau-like levels. In particular, the geometric gauge potential $\mathcal{A}_s$ is a function of $r$, and thus the gap between Landau-like levels is position dependent and is described in Fig.~\ref{Fig3}.

\begin{figure}[htbp]
  \centering
  \includegraphics[width=0.44\textwidth]{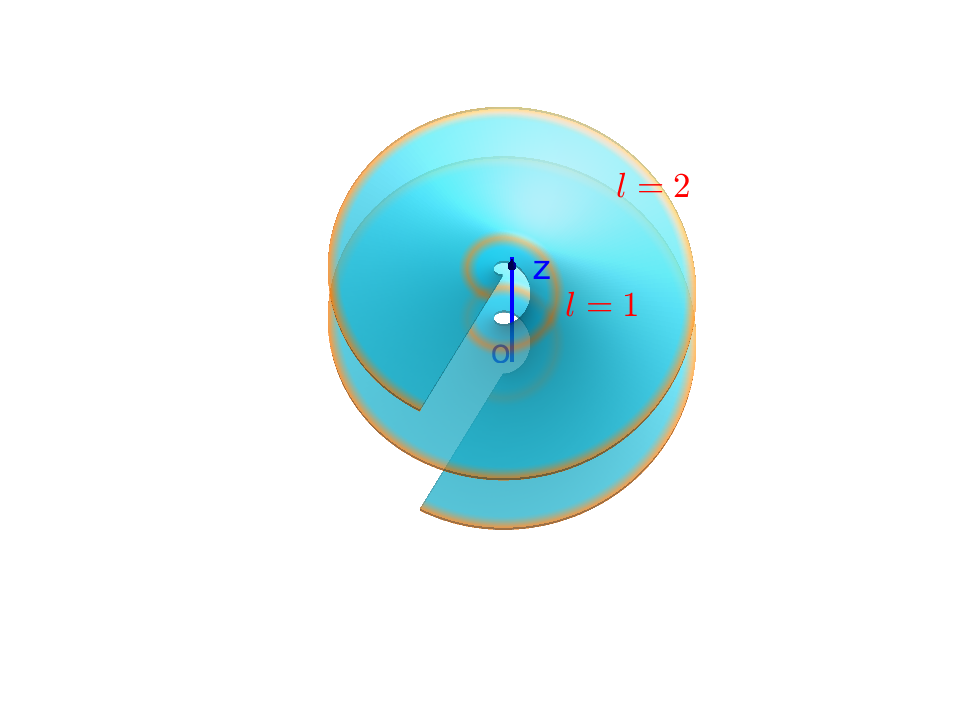}
  \caption{\footnotesize The two channels of $l=1, 2$ for Hall current.}\label{Fig4}
\end{figure}

According to Eqs.~\eqref{SEz},~\eqref{Ez} and~\eqref{EffHamilton}, the $r$ component of the effective Schr\"{o}dinger equation can be simplified as
\begin{equation}\label{SEr}
-\frac{\hbar^2}{2m^*}\frac{d^2}{dr^2}f(r)+U(r)f(r)=Ef(r),
\end{equation}
which represents the motion of the $r$ component with a net potential
\begin{equation}
\begin{split}
U(r)&=E_0+V_g\\
&=-\frac{\hbar^2}{2m^*}\frac{w^2}{\eta^2}\left[\frac{1}{4} -(m+\frac{l}{\sqrt{2}\eta})^2\right].
\end{split}
\end{equation}
Obviously, the Landau-like energy levels are determined by the quantum number of orbital spin $l$, and they are decreasing with increasing $r$ owing to the $r$ dependence of $\eta$. As shown in Fig.~\ref{Fig3}, the two energy levels of $l=1, 2$ cross with $E_F$ at different positions of $r$ simultaneously. There are two channels for Hall current (see Fig.~\ref{Fig4}), as long as the electric field induced by the inhomogeneous distribution of charged particles is strong enough to fully balance $\mathcal{B}_n$. The simultaneous occurrence of multiple Hall conductances is a direct manifestation to the HV.

It is straightforward that $U(r)$ is a sum of two contributions: an attractive part $\frac{1}{4}$ and a repulsive part $(m+\frac{l}{\sqrt{2}\eta})^2$. When $|m+\frac{l}{\sqrt{2}\eta}|<\frac{1}{2}$, $U(r)$ plays the role of an anticentrifugal potential that gathers the particles around the inner rim of $\mathbb{S}^2$. While $|m+\frac{l}{\sqrt{2}\eta}|>\frac{1}{2}$, $U(r)$ is a centrifugal potential that pushes the particle to the outer rim. In the former case, the geometric potential plays a major role, causing the particles to gather to the inner edge. In the latter one, the behavior can be inferred using the uncertainty principle. Localized states tend to appear away from the inner rim. Physically, one may understand the appearance of localized states away from the central axis using the following reasoning: for large $r$ a particle avails more space in $s$ dimension with $ds=\sqrt{1+w^2r^2}dz$, and the momentum and the energy correspondingly decrease.

More specifically, for the contribution of Hall current the two channels have a slight difference that can show some details of the HV. The Hall current contributed by the inner channel is larger than that contributed by the outer one. The current difference can be obtained as
\begin{equation}\label{DeltaCurrent}
\delta I=I_{0}\left(\frac{1}{\sqrt{1+w^2r_1^2}}-\frac{1}{\sqrt{1+w^2r_2^2}}\right),
\end{equation}
where $I_0$ is a current contributed by the charged particles with a unit orbital spin in a flat surface, and $r_{1,2}$ denote the positions of $l=1,2$ channels, respectively. The current difference is initially determined by the different effective velocities of the particle for different $r$ positions, and it will gradually vanish by redistributing particles. For the final equilibrium state, the widening of the channel may compensate the decrease of effective velocity.

\section{Conclusions and Discussions}
As conclusions, we have deduced the effective Hamiltonian describing a particle confined to the helical surface embedded in a 4D Euclidean space by the thin-layer quantization approach. We have found that a geometric gauge potential and a geometric scalar potential appear in the effective Hamiltonian. The geometric gauge potential determines the U(1) structure of the effective Hamiltonian and minimally couples with an intrinsic orbital spin. For charged particles, we further found that the response of orbital spin to torsion contributes to the QHE; the response of the orbital spin to the position dependence of torsion provides the HV. Moreover, the HV is presented as simultaneous occurrence of multiple channels for QHE. Hence, by measuring the Hall current changed by torsion, one can determine the HV.

These imply two results of tantalizing significance. First is that the orbital spin provides a tool to directly probe the HV, a geometrical response, which will significantly extend the research area of QHE. The other is that the torsion can be employed to exhibit a certain QH physics in 4D space. Specifically, the geometrically induced gauge potential can provide a platform to display the quantum physics defined in 4D space. Therefore designing the geometrical and topological structures to improve quantum devices can be employed with multiple applications in quantum computation and information processing remains to be investigated. In other words, the results are helpful to observe and manipulate the HV, a new research area at the frontier of the QH system.

Fortunately, the effective Hamiltonian can be expressed in $\mathbb{R}^3$ to describe the charged particle with an orbital spin confined to a helical surface $\mathbb{S}^2$ really embedded in $\mathbb{R}^3$. Here the orbital spin is an intrinsic degree of freedom, although it originates from the original 4D space $\mathbb{R}^4$. These results provide the possibility for related experiments. Experimentally, the torsion can be provided by disclination~\cite{Lima2012Integer}, screw dislocation~\cite{Filgueiras20152DEG}, or dispiration~\cite{Lima2013Screw}, and the particles with orbital spin can be generated by a spiral phase plate ~\cite{Uchida2010Generation} or by a versatile holographic reconstruction technique~\cite{Verbeeck2010Production}. The geometry of a 2D helical surface can be used to manipulate the phase singularity of the particle with orbital spin~\cite{Silenko2017Manipulating, Lloyd2017Electron}.

\section*{Acknowledges}
This work is jointly supported by the National Nature Science Foundation of China (Grants No. 12075117, No. 51721001, No. 11890702 and No. 11625418, No. 11535005, No. 11690030), National Major State Basic Research and Development of China (Grant No. 2016YFE0129300) and the National Key Research and Development Program of China (Grant No. 2017YFA0303700).

\section*{Appendix A: The effective Hamiltonian for a curved surface embedded in a 4D Euclidean space}
In the Appendix, a 2D curved surface embedded in 4D Euclidean space is considered. For convenience, $\tb{r}(x)$: $\mathbb{M}^2\to\mathbb{R}^4$ is employed to denote the embedding of $\mathbb{M}^2$ in $\mathbb{R}^4$, where $x$ stands for a set of $x^a (a=1, 2)$ denoting the two tangent coordinates of $\mathbb{M}^2$. In the case where $\mathcal{F}$ can be spanned by two tangent vectors $\tb{t}_a=\partial_a\tb{r}$ $(a=1, 2)$ and two normal vectors $\tb{n}_k$ $(k=3, 4)$, the four coordinates are orthonormal to each other. In the immediate neighborhood of $\mathbb{M}^2$, the position vector $\tb{R}$ can be described by $\tb{R}(x,y)=\tb{r}(x)+y_k\tb{n}_k(x)$, where $y_k$ stands for the distance from a point to $\mathbb{M}^2$ along the $\tb{n}_k(x)$ direction. In terms of the definitions $G_{ij}\equiv\partial_i\tb{R}\cdot\partial_j\tb{R}$ $(i, j=1, 2, 3, 4)$ and $g_{ab}\equiv\partial_a\tb{r}\cdot\partial_b\tb{r}$, $G_{ij}$ can be expressed as
\renewcommand\theequation{A1}
\begin{equation}\label{A-MetricFrame}
G_{ij}=
\left (
\begin{array}{ccc}
\gamma_{ab}+y_my_nA_a^{mo}A_{b}^{no} & y_mA_{a}^{lm}\\
y_mA_{b}^{km} & \delta_{kl}
\end{array}
\right ),
\end{equation}
where
\renewcommand\theequation{A2}
\begin{equation}\label{A-GammaMetrix}
\gamma_{ab}=g_{ab}-2y_k\alpha^k_{ab}+y_ky_l\alpha_{ac}^kg^{cd}\alpha_{db}^l.
\end{equation}
Here $\alpha_{ab}^k=\tb{t}_a\cdot\partial_b\tb{n}^k$ is the second fundamental form and $A_a^{lm}=\tb{n}^l\cdot\partial_a\tb{n}^m$ is the normal fundamental form. It is easy to prove that the determinant of $G_{ij}$ is equal to that of $\gamma_{ab}$, $G=\gamma$. According to Eq.~\eqref{A-MetricFrame}, the inverse of $G_{ij}$ can be calculated as
\renewcommand\theequation{A3}
\begin{equation}\label{A-InvMetricFrame}
G^{ij}=
\left (
\begin{array}{ccc}
\gamma^{ab} & \gamma^{ac}y_mA_c^{ml}\\
\gamma^{bc}y_mA_c^{mk} & \delta^{kl}+y_my_nA_c^{km}A_d^{ln}\gamma^{cd}
\end{array}
\right ),
\end{equation}
where $\gamma^{ab}$ is the inverse of $\gamma_{ab}$.

To deduce the effective Hamiltonian for a particle confined to the curved surface $\mathbb{M}^2$ embedded in $\mathbb{R}^4$, we consider a free particle in $\mathbb{R}^4$ that can be described by the Hamiltonian as
\renewcommand\theequation{A4}
\begin{equation}\label{A-FHam}
{\rm{H}}=-\frac{\hbar^2}{2m}\frac{1}{\sqrt{G}}\partial_i\sqrt{G}G^{ij}\partial_j,
\end{equation}
where $(i, j=1, 2, 3, 4)$ describes the four coordinate variables of $\mathbb{R}^4$. As the particle is confined to $\mathbb{M}^2$, in the thin-layer quantization formalism we have to introduce a confining potential~\cite{Costa1981Quantum}, and the Hamiltonian should be replaced by
\renewcommand\theequation{A5}
\begin{equation}\label{A-IHam}
{\rm{H}}=-\frac{\hbar^2}{2m}\frac{1}{\sqrt{G}}\partial_i\sqrt{G}G^{ij}\partial_j +V_{\lambda}(y),
\end{equation}
where $V_{\lambda}(y)$ is the confining potential, and $y$ stands for the coordinates normal to $\mathbb{M}^2$. For convenience, we choose a coordinate frame $\mathcal{F}$~\cite{Jaffe2003Quantum} that consists of two tangent coordinates and two normal coordinates of $\mathbb{M}^2$.

In order to obtain a wave function describing the probability density for a particle moving on $\mathbb{M}^2$, we can in principle introduce a new wave function $|\chi\rangle=|\gamma|^{\frac{1}{4}}|g|^{-\frac{1}{4}}|\psi\rangle$ by rescaling the initial wave function $|\psi\rangle$ with the probability conservation. And then the new wave function can be separated into a tangent component and a normal one analytically with $|\chi\rangle=|\chi_t\rangle |\chi_n\rangle$. Similarly, the Hamiltonian~\eqref{A-IHam} is rescaled by
\renewcommand\theequation{A6}
\begin{equation}\label{A-ResHam0}
\begin{split}
{\rm{H}}^{\prime}&=|\gamma|^{\frac{1}{4}}|g|^{-\frac{1}{4}}{\rm{H}} |g|^{\frac{1}{4}}|\gamma|^{-\frac{1}{4}}\\
& =-\frac{\hbar^2}{2m}|g|^{-\frac{1}{4}}|\gamma|^{-\frac{1}{4}} \partial_i\sqrt{|\gamma|}\gamma^{ij} \partial_j|g|^{\frac{1}{4}}|\gamma|^{-\frac{1}{4}}\\
&\quad +V_{\lambda}(y)\\
&=-\frac{\hbar^2}{2m}|\gamma|^{-\frac{1}{4}}|g|^{-\frac{1}{4}} (\partial_a|\gamma|^{\frac{1}{2}}\gamma^{ab}\partial_b\\
& \quad\quad\quad +\partial_a|\gamma|^{\frac{1}{2}}\gamma^{ac}y_mA_c^{ml}\partial_l\\
&\quad\quad\quad +\partial_k|\gamma|^{\frac{1}{2}}\gamma^{bc}y_mA_c^{mk}\partial_b\\
& \quad\quad\quad +\partial_k|\gamma|^{\frac{1}{2}}y_my_nA_c^{km}A_d^{ln}\gamma^{cd}\partial_l) |g|^{\frac{1}{4}}|\gamma|^{-\frac{1}{4}}\\
&\quad -\frac{\hbar^2}{2m} (|\gamma|^{-\frac{1}{4}}\partial_k|\gamma|^{\frac{1}{2}}\partial_k|\gamma|^{-\frac{1}{4}}) +V_{\lambda}(y).
\end{split}
\end{equation}
In the subspace $\mathbb{N}^2$ spanned by $\tb{n}_3$ and $\tb{n}_4$, an angular momentum operator can be defined by $\hat{L}_{kl}=i(y_l\partial_k-y_k\partial_l)$ $(k, l=3, 4)$. By introducing $D_{a}=\partial_a+\frac{1}{2}iA_a^{kl}\hat{L}_{kl}$ we can compactly rewrite Eq.~\eqref{A-ResHam0} as follows:
\renewcommand\theequation{A7}
\begin{equation}\label{A-ResHam}
\begin{split}
{\rm{H}}^{\prime}=& -\frac{\hbar^2}{2m}[|\gamma|^{-\frac{1}{4}}|g|^{-\frac{1}{4}} (D_a|\gamma|^{\frac{1}{2}}\gamma^{ab}D_b)|g|^{\frac{1}{4}}|\gamma|^{-\frac{1}{4}}\\
& \quad\quad\quad +\frac{3}{16}|\gamma|^{-2}(\partial_k|\gamma|)^2 -\frac{1}{4}|\gamma|^{-1}(\partial_k^2|\gamma|)]\\
& -\frac{\hbar^2}{2m}|\gamma|^{-\frac{1}{4}}|g|^{-\frac{1}{4}} [(\partial_k|\gamma|^{\frac{1}{2}})\gamma^{bc}y_mA_c^{mk}\partial_b\\
& \quad\quad\quad +|\gamma|^{\frac{1}{2}}(\partial_k\gamma^{bc})y_mA_c^{mk}\partial_b\\
& \quad\quad\quad +|\gamma|^{\frac{1}{2}}\gamma^{bc}A_c^{mm}\partial_b] |g|^{\frac{1}{4}}|\gamma|^{-\frac{1}{4}}\\
&-\frac{\hbar^2}{2m}|\gamma|^{-\frac{1}{4}}|g|^{-\frac{1}{4}} [(\partial_k|\gamma|^{\frac{1}{2}})y_my_nA_c^{km}A_d^{ln}\gamma^{cd}\partial_l\\
& \quad\quad\quad +|\gamma|^{\frac{1}{2}}(\partial_k\gamma^{cd})y_my_nA_c^{km}A_d^{ln}\partial_l\\
& \quad\quad\quad +|\gamma|^{\frac{1}{2}}\gamma^{cd}y_nA_c^{mm}A_d^{ln}\partial_l\\
& \quad\quad\quad +|\gamma|^{\frac{1}{2}}\gamma^{cd}y_mA_c^{km}A_d^{nn}\partial_l] |g|^{\frac{1}{4}}|\gamma|^{-\frac{1}{4}}\\
& -\frac{\hbar^2}{2m}\partial_k^2+V_{\lambda}(y).
\end{split}
\end{equation}

For the thin-layer quantization scheme, the aim is to obtain the effective Hamiltonian describing a particle confined to $\mathbb{M}^2$. That is to separate the tangent component of Eq.~\eqref{A-IHam} from the normal one analytically. In other words, the final aim can be accomplished by eliminating the normal part in light of the extreme limit of the confining potential. Therefore, the thin-layer quantization procedure comes down to solve the wave function of ground state of the normal component~\cite{Wang2018Geometric}, which can be simply written as
\renewcommand\theequation{A8}
\begin{equation}\label{A-0Ham}
{\rm{H}_0}=-\frac{\hbar^2}{2m}\frac{1}{\rho}\partial_{\rho}\rho\partial_{\rho} -\frac{\hbar^2}{2m\rho^2}\partial_{\theta}^2+\frac{1}{2}mw_0^2\rho^2,
\end{equation}
where $\rho$ and $\theta$ are two polar coordinate variables that are employed to replace the two coordinate variables $y_3$ and $y_4$ normal to $\mathbb{M}^2$, and the confining potential is selected as  $V_{\lambda}(\rho)=\lim_{w_0\to\infty}\frac{1}{2}mw_0^2\rho^2$, which is invariant under SO(2) transformation. In terms of Eq.~\eqref{A-0Ham}, the wave function of the ground state of $\rho$ can be obtained as
\renewcommand\theequation{A9}
\begin{equation}\label{A-GroundState}
|\chi_{0,l}\rangle=Ae^{il\theta}(B\rho)^{|l|}e^{-B^2\rho^2/2},
\end{equation}
where $A$ is a normalized constant with $A=\sqrt{\frac{2^{|l|+1}B}{\sqrt{\pi}(2|l|-1)!!}}$, $B=\sqrt{\frac{mw_0}{\hbar}}$, and "$!!$" denotes a double factorial. According to the wave function Eq.~\eqref{A-GroundState} and the geometric formula in~\cite{Wang2018Geometric, Wang2018E}, we can express the effective Hamiltonian in the following form
\renewcommand\theequation{A10}
\begin{equation}\label{A-EffHam}
\begin{split}
{\rm{H}_{eff}} & =\langle\chi_{0,l}|{\rm{H}}^{\prime}-{\rm{H}_0}|\chi_{0,l}\rangle_0\\
&=-\frac{\hbar^2}{2m}\langle\chi_{0,l}|[D_aD^a +\frac{3}{16}|\gamma|^{-2}(\partial_k|\gamma|)^2\\
& \quad\quad\quad -\frac{1}{4}|\gamma|^{-1}(\partial_k^2|\gamma|)]|\chi_{0,l}\rangle_0.
\end{split}
\end{equation}
The simple formula is eventually attributed to the SO(2) symmetry of $V_{\lambda}$.

In the presence of $V_{\lambda}$, the subspace $\mathbb{N}^2$ is extremely squeezed. In the case of the sufficiently small size of $\mathbb{N}^2$, we can take
\renewcommand\theequation{A11}
\begin{equation}\label{A-Approx}
\gamma^{ab}=g^{ab}+2y_k\alpha^{ab}_k+3y_ky_l\alpha_k^{cb}\alpha_{lc}^a+O(y^3),
\end{equation}
where $\alpha_k^{ab}$ is the second fundamental form that can be interpreted as the elements of the Weingarten curvature tensor of $\mathbb{M}^2$, and determine the well-known geometric potential~\cite{Costa1981Quantum}. It is worthwhile to notice that the wave function $|\chi_{0,l}\rangle$ in Eq.~\eqref{A-EffHam} is the degenerate nontrivial representation of SO(2) owing to the presence of $V_{\lambda}(\rho)$ in Eq.~\eqref{A-0Ham}. As a result, the angular momentum operator $\hat{L}_{kl}$ is minimally coupled in the covariant derivative operator $D_a$ as a topological charge in $|\chi_{0,l}\rangle$. In $D_a$, the normal fundamental form $A_a^{kl}$ describes a rotation of the plane spanned by $\tb{n}^3$ and $\tb{n}^4$ around the $x^a$ direction, and it is a torsion that plays the role of gauge potential. Therefore the effective Hamiltonian $\rm{H}_{eff}$ describes a particle on a 2D curved surface in the presence of a geometric gauge field and a geometric scalar potential. In other words, the effective dynamics of the particle confined to $\mathbb{M}^2$ is invariant under local SO(2) transformation. As promised, the confining potential $V_{\lambda}$ is invariant under the Abelian $\rm{SO(2)}\simeq\rm{U(1)}$ transformation. Eventually, the U(1) gauge structure of the effective Hamiltonian is from the topological structure of $\mathbb{N}^2$ extrinsic to $\mathbb{M}^2$.

\section*{Appendix B: The geometry of a helical surface embedded in a 4D space}
A helical surface embedded in $\mathbb{R}^4$ is sketched in Fig.~\ref{Fig1} that can be parameterized by
\renewcommand\theequation{B1}
\begin{equation}\label{B-Surface}
\tb{r}(r,z)=(r, z, r\cos wz,r\sin wz),
\end{equation}
where $w$ is the twist angle per unit length along the $z$ axis (see Fig.~\ref{Fig1}), and $w=2\pi/d$, wherein $d$ is the screw pitch, $r$ is the distance to the central axis, $r\in[r_{in},r_{out}]$, and the width of the strip is defined by $D=r_{out}-r_{in}$. Here the $r$ and $z$ coordinates have been mapped onto the helical surface $\mathbb{S}^2$ embedded in $\mathbb{R}^4$. For the convenience of description, an adapted frame $\mathcal{F}$ is employed to describe the neighbourhood subspace of $\mathbb{S}^2$ that is spanned by two tangent basis vectors $\tb{t}_r$ and $\tb{t}_z$ and two normal basis vectors $\tb{n}_3$ and $\tb{n}_4$ of $\mathbb{S}^2$. With Eq.~\eqref{B-Surface}, the local orthogonal four basis vectors can be calculated as
\renewcommand\theequation{B2}
\begin{equation}\label{B-AdaptedFrame}
\begin{split}
& \tb{t}_r=(1, 0, \cos wz, \sin wz),\\
& \tb{t}_z=(0, 1, -wr\sin wz, wr\cos wz),\\
& \tb{n}_3=\frac{1}{\sqrt{2}}(-1, 0, \cos wz, \sin wz),\\
& \tb{n}_4=\frac{1}{\eta}(0, -wr, -\sin wz, \cos wz),
\end{split}
\end{equation}
where $\eta=\sqrt{1+w^2r^2}$. In terms of Eq.~\eqref{B-AdaptedFrame} and the definitions $g_{ab}\equiv\partial_a\tb{r}\cdot\partial_b\tb{r}$, $\alpha_{ab}^k=\tb{t}_a\cdot\partial_b\tb{n}^k$ and $A_a^{lm}=\tb{n}^l\cdot\partial_a\tb{n}^m$, $(a,b=r,z)$ and $(k, l, m=3, 4)$, we can obtain the first fundamental form of the metric tensor $g_{ab}$ as
\renewcommand\theequation{B3}
\begin{equation}\label{B-MetricS}
g_{ab}=\left (
\begin{array}{ccc}
2 & 0\\
 0 & \eta^2
 \end{array}
\right ),
\end{equation}
the second fundamental form of the Weingarten curvature tensor $\alpha_{ab}^3$ and $\alpha_{ab}^4$ as
\renewcommand\theequation{B4}
\begin{equation}\label{B-WCurvature}
\alpha_{ab}^3=\left (
\begin{array}{ccc}
0 & 0\\
 0 & \frac{w^2 r}{\sqrt{2}},
 \end{array}
\right ),\quad \alpha_{ab}^4=\left (
\begin{array}{ccc}
0 & -\frac{w}{\eta}\\
-\frac{w}{\eta} & 0,
\end{array}
\right )
\end{equation}
and the normal fundamental form of the connections $A_1^{ij}$ and $A_2^{ij}$ as
\renewcommand\theequation{B5}
\begin{equation}\label{B-NormConnect}
A_1^{ij}=\left (
\begin{array}{ccc}
0 & 0\\
 0 & 0,
 \end{array}
\right ),\quad A_2^{ij}=\left (
\begin{array}{ccc}
0 & -\frac{w}{\sqrt{2}\eta}\\
\frac{w}{\sqrt{2}\eta} & 0,
\end{array}
\right ).
\end{equation}
From Eq.~\eqref{B-MetricS} it is easy to obtain the determinant of $g_{ab}$ as
\renewcommand\theequation{B6}
\begin{equation}\label{g}
g=2\eta^2.
\end{equation}
According to Eqs.~\eqref{A-GammaMetrix}, ~\eqref{B-MetricS}, ~\eqref{B-WCurvature} and ~\eqref{B-NormConnect}, the matrix $\gamma_{ab}$ can be deduced as
\begin{widetext}
\renewcommand\theequation{B7}
\begin{equation}\label{B-GammaM}
\gamma_{ab}=\left (
\begin{array}{ccc}
2+\frac{w^2}{\eta^4}q_4^2 & -\frac{2w}{\eta}q_4-\frac{w^3 r}{\sqrt{2}\eta^3}q_3q_4\\
-\frac{2w}{\eta}q_4-\frac{w^3r}{\sqrt{2}\eta}q_3q_4 & \eta^2+\sqrt{2}w^2 r q_3+\frac{w^4r^2}{2\eta^2}q_3^2+\frac{w^2}{2\eta^2}q_4^2
\end{array}
\right ).
\end{equation}
Subsequently, the determinant and the inverse of $\gamma_{ab}$ are calculated as
\renewcommand\theequation{B8}
\begin{equation}\label{B-GammaD}
\gamma=2\eta^2+2\sqrt{2}w^2rq_3+\frac{w^4r^2}{\eta^2}q_3^2 -\frac{2w^2}{\eta^2}q_4^2
-\frac{\sqrt{2}w^4r}{\eta^4}q_3 q_4^2+\frac{w^4}{2\eta^6}q_4^4,
\end{equation}
and
\renewcommand\theequation{B9}
\begin{equation}\label{B-InvGammaM}
\gamma^{ab}=\frac{1}{D}\left (
\begin{array}{ccc}
2\eta^8+2\sqrt{2}\eta^6w^2rq_3+\eta^4w^4r^2q_3^2+\eta^4w^2q_4^2 & 4\eta^5w q_4+\sqrt{2}\eta^3w^3 rq_3q_4\\
4\eta^5w q_4+\sqrt{2}\eta^3w^3 rq_3q_4 & 4\eta^6+2\eta^2w^2 q_4^2,
\end{array}
\right )
\end{equation}
respectively, where $D=4\eta^8 +4\sqrt{2}\eta^6w^2rq_3 +2\eta^4w^4r^2q_3^2-4\eta^4w^2q_4^2 -2\sqrt{2}\eta^2w^4rq_3q_4^2 +w^4q_4^4$.

In the thin-layer quantization scheme, it is important that the confining potential with SO(2) invariance is introduced to reduce the normal dimensions. The normal dimensions $q_3$ and $q_4$ are confined in infinitesimal intervals, in which the wave function satisfies Gaussian distribution. By virtue of the infinitesimal values of $q_3$ and $q_4$ and the simple form of $|\chi_{0, l}\rangle$ Eq.~\eqref{A-GroundState} and the geometric formula Eq.~\eqref{A-EffHam}, we can use the following approximate expressions
\renewcommand\theequation{B10}
\begin{equation}\label{B-Gamma_appr}
\gamma^{1/2}\approx \sqrt{2}\eta+\frac{w^2r}{\eta}q_3 -\frac{w^2}{\sqrt{2}\eta^3}q_4^2,
\end{equation}
\renewcommand\theequation{B11}
\begin{equation}\label{B-SqrtFact}
\begin{split}
\gamma^{1/4}g^{-1/4}\approx & 1+\frac{\sqrt{2}}{4\eta^2}w^2rq_3\\
& -\frac{1}{16\eta^4}w^4r^2q_3^2 -\frac{1}{4\eta^4}w^2q_4^2,
\end{split}
\end{equation}
and
\renewcommand\theequation{B12}
\begin{equation}\label{B-ISqrtFact}
\begin{split}
\gamma^{-1/4}g^{1/4}\approx & 1-\frac{3\sqrt{2}}{8\eta^2}w^2rq_3\\
& +\frac{21}{64\eta^4}w^4r^2q_3^2+\frac{3}{8\eta^4}w^2q_4^2,
\end{split}
\end{equation}
respectively.

In terms of Eq.~\eqref{A-GroundState}, Eq.~\eqref{A-EffHam} and the above approximations, the effective Hamiltonian describing a particle confined to $\mathbb{M}^2$ can be obtained as
\begin{equation}\label{HSEffHam0}
{\rm{H}}_{\rm{eff}}=-\frac{\hbar^2}{2m}\frac{1}{\eta^2}(\partial_z+il\tau)^2 -\frac{\hbar^2}{4m}\partial_r^2 -\frac{\hbar^2}{4m}\tau^2,
\end{equation}
with $\tau=\frac{w}{\sqrt{2(1+w^2r^2)}}$ being the torsion of $\mathbb{S}^2$.
\end{widetext}

\bibliographystyle{apsrev4-1}
\bibliography{QHEG}

\begin{thebibliography}{47}%
\makeatletter
\providecommand \@ifxundefined [1]{%
 \@ifx{#1\undefined}
}%
\providecommand \@ifnum [1]{%
 \ifnum #1\expandafter \@firstoftwo
 \else \expandafter \@secondoftwo
 \fi
}%
\providecommand \@ifx [1]{%
 \ifx #1\expandafter \@firstoftwo
 \else \expandafter \@secondoftwo
 \fi
}%
\providecommand \natexlab [1]{#1}%
\providecommand \enquote  [1]{``#1''}%
\providecommand \bibnamefont  [1]{#1}%
\providecommand \bibfnamefont [1]{#1}%
\providecommand \citenamefont [1]{#1}%
\providecommand \href@noop [0]{\@secondoftwo}%
\providecommand \href [0]{\begingroup \@sanitize@url \@href}%
\providecommand \@href[1]{\@@startlink{#1}\@@href}%
\providecommand \@@href[1]{\endgroup#1\@@endlink}%
\providecommand \@sanitize@url [0]{\catcode `\\12\catcode `\$12\catcode
  `\&12\catcode `\#12\catcode `\^12\catcode `\_12\catcode `\%12\relax}%
\providecommand \@@startlink[1]{}%
\providecommand \@@endlink[0]{}%
\providecommand \url  [0]{\begingroup\@sanitize@url \@url }%
\providecommand \@url [1]{\endgroup\@href {#1}{\urlprefix }}%
\providecommand \urlprefix  [0]{URL }%
\providecommand \Eprint [0]{\href }%
\providecommand \doibase [0]{http://dx.doi.org/}%
\providecommand \selectlanguage [0]{\@gobble}%
\providecommand \bibinfo  [0]{\@secondoftwo}%
\providecommand \bibfield  [0]{\@secondoftwo}%
\providecommand \translation [1]{[#1]}%
\providecommand \BibitemOpen [0]{}%
\providecommand \bibitemStop [0]{}%
\providecommand \bibitemNoStop [0]{.\EOS\space}%
\providecommand \EOS [0]{\spacefactor3000\relax}%
\providecommand \BibitemShut  [1]{\csname bibitem#1\endcsname}%
\let\auto@bib@innerbib\@empty
\bibitem [{\citenamefont {Klitzing}\ \emph {et~al.}(1980)\citenamefont
  {Klitzing}, \citenamefont {Dorda},\ and\ \citenamefont
  {Pepper}}]{Klitzing1980New}%
  \BibitemOpen
  \bibfield  {author} {\bibinfo {author} {\bibfnamefont {K.~v.}\ \bibnamefont
  {Klitzing}}, \bibinfo {author} {\bibfnamefont {G.}~\bibnamefont {Dorda}}, \
  and\ \bibinfo {author} {\bibfnamefont {M.}~\bibnamefont {Pepper}},\ }\href
  {\doibase 10.1103/PhysRevLett.45.494} {\bibfield  {journal} {\bibinfo
  {journal} {Phys. Rev. Lett.}\ }\textbf {\bibinfo {volume} {45}},\ \bibinfo
  {pages} {494} (\bibinfo {year} {1980})}\BibitemShut {NoStop}%
\bibitem [{\citenamefont {Tsui}\ \emph {et~al.}(1982)\citenamefont {Tsui},
  \citenamefont {Stormer},\ and\ \citenamefont {Gossard}}]{Tsui1982Two}%
  \BibitemOpen
  \bibfield  {author} {\bibinfo {author} {\bibfnamefont {D.~C.}\ \bibnamefont
  {Tsui}}, \bibinfo {author} {\bibfnamefont {H.~L.}\ \bibnamefont {Stormer}}, \
  and\ \bibinfo {author} {\bibfnamefont {A.~C.}\ \bibnamefont {Gossard}},\
  }\href {\doibase 10.1103/PhysRevLett.48.1559} {\bibfield  {journal} {\bibinfo
   {journal} {Phys. Rev. Lett.}\ }\textbf {\bibinfo {volume} {48}},\ \bibinfo
  {pages} {1559} (\bibinfo {year} {1982})}\BibitemShut {NoStop}%
\bibitem [{\citenamefont {Thouless}\ \emph {et~al.}(1982)\citenamefont
  {Thouless}, \citenamefont {Kohmoto}, \citenamefont {Nightingale},\ and\
  \citenamefont {den Nijs}}]{Thouless1982Quantized}%
  \BibitemOpen
  \bibfield  {author} {\bibinfo {author} {\bibfnamefont {D.~J.}\ \bibnamefont
  {Thouless}}, \bibinfo {author} {\bibfnamefont {M.}~\bibnamefont {Kohmoto}},
  \bibinfo {author} {\bibfnamefont {M.~P.}\ \bibnamefont {Nightingale}}, \ and\
  \bibinfo {author} {\bibfnamefont {M.}~\bibnamefont {den Nijs}},\ }\href
  {\doibase 10.1103/PhysRevLett.49.405} {\bibfield  {journal} {\bibinfo
  {journal} {Phys. Rev. Lett.}\ }\textbf {\bibinfo {volume} {49}},\ \bibinfo
  {pages} {405} (\bibinfo {year} {1982})}\BibitemShut {NoStop}%
\bibitem [{\citenamefont {Avron}\ \emph {et~al.}(1995)\citenamefont {Avron},
  \citenamefont {Seiler},\ and\ \citenamefont {Zograf}}]{Avron1995Viscosity}%
  \BibitemOpen
  \bibfield  {author} {\bibinfo {author} {\bibfnamefont {J.~E.}\ \bibnamefont
  {Avron}}, \bibinfo {author} {\bibfnamefont {R.}~\bibnamefont {Seiler}}, \
  and\ \bibinfo {author} {\bibfnamefont {P.~G.}\ \bibnamefont {Zograf}},\
  }\href {\doibase 10.1103/PhysRevLett.75.697} {\bibfield  {journal} {\bibinfo
  {journal} {Phys. Rev. Lett.}\ }\textbf {\bibinfo {volume} {75}},\ \bibinfo
  {pages} {697} (\bibinfo {year} {1995})}\BibitemShut {NoStop}%
\bibitem [{\citenamefont {Can}\ \emph {et~al.}(2014)\citenamefont {Can},
  \citenamefont {Laskin},\ and\ \citenamefont {Wiegmann}}]{Can2014Fractional}%
  \BibitemOpen
  \bibfield  {author} {\bibinfo {author} {\bibfnamefont {T.}~\bibnamefont
  {Can}}, \bibinfo {author} {\bibfnamefont {M.}~\bibnamefont {Laskin}}, \ and\
  \bibinfo {author} {\bibfnamefont {P.}~\bibnamefont {Wiegmann}},\ }\href
  {\doibase 10.1103/PhysRevLett.113.046803} {\bibfield  {journal} {\bibinfo
  {journal} {Phys. Rev. Lett.}\ }\textbf {\bibinfo {volume} {113}},\ \bibinfo
  {pages} {046803} (\bibinfo {year} {2014})}\BibitemShut {NoStop}%
\bibitem [{\citenamefont {Wiegmann}(2018)}]{Wiegmann2018Inner}%
  \BibitemOpen
  \bibfield  {author} {\bibinfo {author} {\bibfnamefont {P.}~\bibnamefont
  {Wiegmann}},\ }\href {\doibase 10.1103/PhysRevLett.120.086601} {\bibfield
  {journal} {\bibinfo  {journal} {Phys. Rev. Lett.}\ }\textbf {\bibinfo
  {volume} {120}},\ \bibinfo {pages} {086601} (\bibinfo {year}
  {2018})}\BibitemShut {NoStop}%
\bibitem [{\citenamefont {Gromov}\ \emph {et~al.}(2015)\citenamefont {Gromov},
  \citenamefont {Cho}, \citenamefont {You}, \citenamefont {Abanov},\ and\
  \citenamefont {Fradkin}}]{Fradkin2015Framing}%
  \BibitemOpen
  \bibfield  {author} {\bibinfo {author} {\bibfnamefont {A.}~\bibnamefont
  {Gromov}}, \bibinfo {author} {\bibfnamefont {G.~Y.}\ \bibnamefont {Cho}},
  \bibinfo {author} {\bibfnamefont {Y.}~\bibnamefont {You}}, \bibinfo {author}
  {\bibfnamefont {A.~G.}\ \bibnamefont {Abanov}}, \ and\ \bibinfo {author}
  {\bibfnamefont {E.}~\bibnamefont {Fradkin}},\ }\href {\doibase
  10.1103/PhysRevLett.114.016805} {\bibfield  {journal} {\bibinfo  {journal}
  {Phys. Rev. Lett.}\ }\textbf {\bibinfo {volume} {114}},\ \bibinfo {pages}
  {016805} (\bibinfo {year} {2015})}\BibitemShut {NoStop}%
\bibitem [{\citenamefont {Hoyos}\ and\ \citenamefont
  {Son}(2012)}]{Hoyos2012Hall}%
  \BibitemOpen
  \bibfield  {author} {\bibinfo {author} {\bibfnamefont {C.}~\bibnamefont
  {Hoyos}}\ and\ \bibinfo {author} {\bibfnamefont {D.~T.}\ \bibnamefont
  {Son}},\ }\href {\doibase 10.1103/PhysRevLett.108.066805} {\bibfield
  {journal} {\bibinfo  {journal} {Phys. Rev. Lett.}\ }\textbf {\bibinfo
  {volume} {108}},\ \bibinfo {pages} {066805} (\bibinfo {year}
  {2012})}\BibitemShut {NoStop}%
\bibitem [{\citenamefont {Berdyugin}\ \emph {et~al.}(2019)\citenamefont
  {Berdyugin}, \citenamefont {Xu}, \citenamefont {Pellegrino}, \citenamefont
  {Krishna~Kumar}, \citenamefont {Principi}, \citenamefont {Torre},
  \citenamefont {Ben~Shalom}, \citenamefont {Taniguchi}, \citenamefont
  {Watanabe}, \citenamefont {Grigorieva}, \citenamefont {Polini}, \citenamefont
  {Geim},\ and\ \citenamefont {Bandurin}}]{Berdyugin2019Measuring}%
  \BibitemOpen
  \bibfield  {author} {\bibinfo {author} {\bibfnamefont {A.~I.}\ \bibnamefont
  {Berdyugin}}, \bibinfo {author} {\bibfnamefont {S.~G.}\ \bibnamefont {Xu}},
  \bibinfo {author} {\bibfnamefont {F.~M.~D.}\ \bibnamefont {Pellegrino}},
  \bibinfo {author} {\bibfnamefont {R.}~\bibnamefont {Krishna~Kumar}}, \bibinfo
  {author} {\bibfnamefont {A.}~\bibnamefont {Principi}}, \bibinfo {author}
  {\bibfnamefont {I.}~\bibnamefont {Torre}}, \bibinfo {author} {\bibfnamefont
  {M.}~\bibnamefont {Ben~Shalom}}, \bibinfo {author} {\bibfnamefont
  {T.}~\bibnamefont {Taniguchi}}, \bibinfo {author} {\bibfnamefont
  {K.}~\bibnamefont {Watanabe}}, \bibinfo {author} {\bibfnamefont {I.~V.}\
  \bibnamefont {Grigorieva}}, \bibinfo {author} {\bibfnamefont
  {M.}~\bibnamefont {Polini}}, \bibinfo {author} {\bibfnamefont {A.~K.}\
  \bibnamefont {Geim}}, \ and\ \bibinfo {author} {\bibfnamefont {D.~A.}\
  \bibnamefont {Bandurin}},\ }\href {\doibase 10.1126/science.aau0685}
  {\bibfield  {journal} {\bibinfo  {journal} {Science}\ }\textbf {\bibinfo
  {volume} {364}},\ \bibinfo {pages} {162} (\bibinfo {year}
  {2019})}\BibitemShut {NoStop}%
\bibitem [{\citenamefont {Haldane}(1988)}]{Haldane1988Model}%
  \BibitemOpen
  \bibfield  {author} {\bibinfo {author} {\bibfnamefont {F.~D.~M.}\
  \bibnamefont {Haldane}},\ }\href {\doibase 10.1103/PhysRevLett.61.2015}
  {\bibfield  {journal} {\bibinfo  {journal} {Phys. Rev. Lett.}\ }\textbf
  {\bibinfo {volume} {61}},\ \bibinfo {pages} {2015} (\bibinfo {year}
  {1988})}\BibitemShut {NoStop}%
\bibitem [{\citenamefont {Wen}(1991)}]{Wen1991Mean}%
  \BibitemOpen
  \bibfield  {author} {\bibinfo {author} {\bibfnamefont {X.~G.}\ \bibnamefont
  {Wen}},\ }\href {\doibase 10.1103/PhysRevB.44.2664} {\bibfield  {journal}
  {\bibinfo  {journal} {Phys. Rev. B}\ }\textbf {\bibinfo {volume} {44}},\
  \bibinfo {pages} {2664} (\bibinfo {year} {1991})}\BibitemShut {NoStop}%
\bibitem [{\citenamefont {Brand{\~{a}}o}\ \emph {et~al.}(2017)\citenamefont
  {Brand{\~{a}}o}, \citenamefont {Filgueiras}, \citenamefont {Rojas},\ and\
  \citenamefont {Moraes}}]{Brandao2017Inertial}%
  \BibitemOpen
  \bibfield  {author} {\bibinfo {author} {\bibfnamefont {J.}~\bibnamefont
  {Brand{\~{a}}o}}, \bibinfo {author} {\bibfnamefont {C.}~\bibnamefont
  {Filgueiras}}, \bibinfo {author} {\bibfnamefont {M.}~\bibnamefont {Rojas}}, \
  and\ \bibinfo {author} {\bibfnamefont {F.}~\bibnamefont {Moraes}},\ }\href
  {\doibase 10.1088/2399-6528/aa8aa3} {\bibfield  {journal} {\bibinfo
  {journal} {J. Phys. Commun.}\ }\textbf {\bibinfo {volume} {1}},\ \bibinfo
  {pages} {035004} (\bibinfo {year} {2017})}\BibitemShut {NoStop}%
\bibitem [{\citenamefont {Nagaosa}\ \emph {et~al.}(2010)\citenamefont
  {Nagaosa}, \citenamefont {Sinova}, \citenamefont {Onoda}, \citenamefont
  {MacDonald},\ and\ \citenamefont {Ong}}]{Nagaosa2010Anomalous}%
  \BibitemOpen
  \bibfield  {author} {\bibinfo {author} {\bibfnamefont {N.}~\bibnamefont
  {Nagaosa}}, \bibinfo {author} {\bibfnamefont {J.}~\bibnamefont {Sinova}},
  \bibinfo {author} {\bibfnamefont {S.}~\bibnamefont {Onoda}}, \bibinfo
  {author} {\bibfnamefont {A.~H.}\ \bibnamefont {MacDonald}}, \ and\ \bibinfo
  {author} {\bibfnamefont {N.~P.}\ \bibnamefont {Ong}},\ }\href {\doibase
  10.1103/RevModPhys.82.1539} {\bibfield  {journal} {\bibinfo  {journal} {Rev.
  Mod. Phys.}\ }\textbf {\bibinfo {volume} {82}},\ \bibinfo {pages} {1539}
  (\bibinfo {year} {2010})}\BibitemShut {NoStop}%
\bibitem [{\citenamefont {Hasan}\ and\ \citenamefont
  {Kane}(2010)}]{Hasan2010Topological}%
  \BibitemOpen
  \bibfield  {author} {\bibinfo {author} {\bibfnamefont {M.~Z.}\ \bibnamefont
  {Hasan}}\ and\ \bibinfo {author} {\bibfnamefont {C.~L.}\ \bibnamefont
  {Kane}},\ }\href {\doibase 10.1103/RevModPhys.82.3045} {\bibfield  {journal}
  {\bibinfo  {journal} {Rev. Mod. Phys.}\ }\textbf {\bibinfo {volume} {82}},\
  \bibinfo {pages} {3045} (\bibinfo {year} {2010})}\BibitemShut {NoStop}%
\bibitem [{\citenamefont {Chang}\ \emph {et~al.}(2013)\citenamefont {Chang},
  \citenamefont {Zhang}, \citenamefont {Feng}, \citenamefont {Shen},
  \citenamefont {Zhang}, \citenamefont {Guo}, \citenamefont {Li}, \citenamefont
  {Ou}, \citenamefont {Wei}, \citenamefont {Wang}, \citenamefont {Ji},
  \citenamefont {Feng}, \citenamefont {Ji}, \citenamefont {Chen}, \citenamefont
  {Jia}, \citenamefont {Dai}, \citenamefont {Fang}, \citenamefont {Zhang},
  \citenamefont {He}, \citenamefont {Wang}, \citenamefont {Lu}, \citenamefont
  {Ma},\ and\ \citenamefont {Xue}}]{XueQK2013Experimental}%
  \BibitemOpen
  \bibfield  {author} {\bibinfo {author} {\bibfnamefont {C.-Z.}\ \bibnamefont
  {Chang}}, \bibinfo {author} {\bibfnamefont {J.}~\bibnamefont {Zhang}},
  \bibinfo {author} {\bibfnamefont {X.}~\bibnamefont {Feng}}, \bibinfo {author}
  {\bibfnamefont {J.}~\bibnamefont {Shen}}, \bibinfo {author} {\bibfnamefont
  {Z.}~\bibnamefont {Zhang}}, \bibinfo {author} {\bibfnamefont
  {M.}~\bibnamefont {Guo}}, \bibinfo {author} {\bibfnamefont {K.}~\bibnamefont
  {Li}}, \bibinfo {author} {\bibfnamefont {Y.}~\bibnamefont {Ou}}, \bibinfo
  {author} {\bibfnamefont {P.}~\bibnamefont {Wei}}, \bibinfo {author}
  {\bibfnamefont {L.-L.}\ \bibnamefont {Wang}}, \bibinfo {author}
  {\bibfnamefont {Z.-Q.}\ \bibnamefont {Ji}}, \bibinfo {author} {\bibfnamefont
  {Y.}~\bibnamefont {Feng}}, \bibinfo {author} {\bibfnamefont {S.}~\bibnamefont
  {Ji}}, \bibinfo {author} {\bibfnamefont {X.}~\bibnamefont {Chen}}, \bibinfo
  {author} {\bibfnamefont {J.}~\bibnamefont {Jia}}, \bibinfo {author}
  {\bibfnamefont {X.}~\bibnamefont {Dai}}, \bibinfo {author} {\bibfnamefont
  {Z.}~\bibnamefont {Fang}}, \bibinfo {author} {\bibfnamefont {S.-C.}\
  \bibnamefont {Zhang}}, \bibinfo {author} {\bibfnamefont {K.}~\bibnamefont
  {He}}, \bibinfo {author} {\bibfnamefont {Y.}~\bibnamefont {Wang}}, \bibinfo
  {author} {\bibfnamefont {L.}~\bibnamefont {Lu}}, \bibinfo {author}
  {\bibfnamefont {X.-C.}\ \bibnamefont {Ma}}, \ and\ \bibinfo {author}
  {\bibfnamefont {Q.-K.}\ \bibnamefont {Xue}},\ }\href {\doibase
  10.1126/science.1234414} {\bibfield  {journal} {\bibinfo  {journal}
  {Science}\ }\textbf {\bibinfo {volume} {340}},\ \bibinfo {pages} {167}
  (\bibinfo {year} {2013})}\BibitemShut {NoStop}%
\bibitem [{\citenamefont {Liu}\ \emph {et~al.}(2016)\citenamefont {Liu},
  \citenamefont {Zhang},\ and\ \citenamefont {Qi}}]{QiXL2016The}%
  \BibitemOpen
  \bibfield  {author} {\bibinfo {author} {\bibfnamefont {C.-X.}\ \bibnamefont
  {Liu}}, \bibinfo {author} {\bibfnamefont {S.-C.}\ \bibnamefont {Zhang}}, \
  and\ \bibinfo {author} {\bibfnamefont {X.-L.}\ \bibnamefont {Qi}},\ }\href
  {\doibase 10.1146/annurev-conmatphys-031115-011417} {\bibfield  {journal}
  {\bibinfo  {journal} {Annu. Rev. Condens. Matter Phys.}\ }\textbf {\bibinfo
  {volume} {7}},\ \bibinfo {pages} {301} (\bibinfo {year} {2016})}\BibitemShut
  {NoStop}%
\bibitem [{\citenamefont {Gromov}\ and\ \citenamefont
  {Abanov}(2014)}]{Gromov2014Density}%
  \BibitemOpen
  \bibfield  {author} {\bibinfo {author} {\bibfnamefont {A.}~\bibnamefont
  {Gromov}}\ and\ \bibinfo {author} {\bibfnamefont {A.~G.}\ \bibnamefont
  {Abanov}},\ }\href {\doibase 10.1103/PhysRevLett.113.266802} {\bibfield
  {journal} {\bibinfo  {journal} {Phys. Rev. Lett.}\ }\textbf {\bibinfo
  {volume} {113}},\ \bibinfo {pages} {266802} (\bibinfo {year}
  {2014})}\BibitemShut {NoStop}%
\bibitem [{\citenamefont {Fujii}\ \emph {et~al.}(1997)\citenamefont {Fujii},
  \citenamefont {Ogawa},\ and\ \citenamefont
  {Uchiyama}}]{Fujii1997Geometrically}%
  \BibitemOpen
  \bibfield  {author} {\bibinfo {author} {\bibfnamefont {K.}~\bibnamefont
  {Fujii}}, \bibinfo {author} {\bibfnamefont {N.}~\bibnamefont {Ogawa}}, \ and\
  \bibinfo {author} {\bibfnamefont {S.}~\bibnamefont {Uchiyama}},\ }\href
  {\doibase 10.1142/S0217751X97002814} {\bibfield  {journal} {\bibinfo
  {journal} {Int. J. Mod. Phys. A}\ }\textbf {\bibinfo {volume} {12}},\
  \bibinfo {pages} {5235} (\bibinfo {year} {1997})}\BibitemShut {NoStop}%
\bibitem [{\citenamefont {Schuster}\ and\ \citenamefont
  {Jaffe}(2003)}]{Jaffe2003Quantum}%
  \BibitemOpen
  \bibfield  {author} {\bibinfo {author} {\bibfnamefont {P.}~\bibnamefont
  {Schuster}}\ and\ \bibinfo {author} {\bibfnamefont {R.}~\bibnamefont
  {Jaffe}},\ }\href {\doibase http://dx.doi.org/10.1016/S0003-4916(03)00080-0}
  {\bibfield  {journal} {\bibinfo  {journal} {Ann. Phys.}\ }\textbf {\bibinfo
  {volume} {307}},\ \bibinfo {pages} {132 } (\bibinfo {year}
  {2003})}\BibitemShut {NoStop}%
\bibitem [{\citenamefont {Wang}\ \emph
  {et~al.}(2018{\natexlab{a}})\citenamefont {Wang}, \citenamefont {Lai},
  \citenamefont {Wang}, \citenamefont {Zong},\ and\ \citenamefont
  {Chen}}]{Wang2018Geometric}%
  \BibitemOpen
  \bibfield  {author} {\bibinfo {author} {\bibfnamefont {Y.-L.}\ \bibnamefont
  {Wang}}, \bibinfo {author} {\bibfnamefont {M.-Y.}\ \bibnamefont {Lai}},
  \bibinfo {author} {\bibfnamefont {F.}~\bibnamefont {Wang}}, \bibinfo {author}
  {\bibfnamefont {H.-S.}\ \bibnamefont {Zong}}, \ and\ \bibinfo {author}
  {\bibfnamefont {Y.-F.}\ \bibnamefont {Chen}},\ }\href {\doibase
  10.1103/PhysRevA.97.042108} {\bibfield  {journal} {\bibinfo  {journal} {Phys.
  Rev. A}\ }\textbf {\bibinfo {volume} {97}},\ \bibinfo {pages} {042108}
  (\bibinfo {year} {2018}{\natexlab{a}})}\BibitemShut {NoStop}%
\bibitem [{\citenamefont {Wang}\ \emph
  {et~al.}(2018{\natexlab{b}})\citenamefont {Wang}, \citenamefont {Lai},
  \citenamefont {Wang}, \citenamefont {Zong},\ and\ \citenamefont
  {Chen}}]{Wang2018E}%
  \BibitemOpen
  \bibfield  {author} {\bibinfo {author} {\bibfnamefont {Y.-L.}\ \bibnamefont
  {Wang}}, \bibinfo {author} {\bibfnamefont {M.-Y.}\ \bibnamefont {Lai}},
  \bibinfo {author} {\bibfnamefont {F.}~\bibnamefont {Wang}}, \bibinfo {author}
  {\bibfnamefont {H.-S.}\ \bibnamefont {Zong}}, \ and\ \bibinfo {author}
  {\bibfnamefont {Y.-F.}\ \bibnamefont {Chen}},\ }\href {\doibase
  10.1103/PhysRevA.97.069904} {\bibfield  {journal} {\bibinfo  {journal} {Phys.
  Rev. A}\ }\textbf {\bibinfo {volume} {97}},\ \bibinfo {pages} {069904(E)}
  (\bibinfo {year} {2018}{\natexlab{b}})}\BibitemShut {NoStop}%
\bibitem [{\citenamefont {Chen}\ \emph {et~al.}(2008)\citenamefont {Chen},
  \citenamefont {L\"u}, \citenamefont {Sun}, \citenamefont {Wang},\ and\
  \citenamefont {Goldman}}]{Chen2008Spin}%
  \BibitemOpen
  \bibfield  {author} {\bibinfo {author} {\bibfnamefont {X.-S.}\ \bibnamefont
  {Chen}}, \bibinfo {author} {\bibfnamefont {X.-F.}\ \bibnamefont {L\"u}},
  \bibinfo {author} {\bibfnamefont {W.-M.}\ \bibnamefont {Sun}}, \bibinfo
  {author} {\bibfnamefont {F.}~\bibnamefont {Wang}}, \ and\ \bibinfo {author}
  {\bibfnamefont {T.}~\bibnamefont {Goldman}},\ }\href {\doibase
  10.1103/PhysRevLett.100.232002} {\bibfield  {journal} {\bibinfo  {journal}
  {Phys. Rev. Lett.}\ }\textbf {\bibinfo {volume} {100}},\ \bibinfo {pages}
  {232002} (\bibinfo {year} {2008})}\BibitemShut {NoStop}%
\bibitem [{\citenamefont {Chen}\ \emph {et~al.}(2009)\citenamefont {Chen},
  \citenamefont {Sun}, \citenamefont {L\"u}, \citenamefont {Wang},\ and\
  \citenamefont {Goldman}}]{Chen2009Do}%
  \BibitemOpen
  \bibfield  {author} {\bibinfo {author} {\bibfnamefont {X.-S.}\ \bibnamefont
  {Chen}}, \bibinfo {author} {\bibfnamefont {W.-M.}\ \bibnamefont {Sun}},
  \bibinfo {author} {\bibfnamefont {X.-F.}\ \bibnamefont {L\"u}}, \bibinfo
  {author} {\bibfnamefont {F.}~\bibnamefont {Wang}}, \ and\ \bibinfo {author}
  {\bibfnamefont {T.}~\bibnamefont {Goldman}},\ }\href {\doibase
  10.1103/PhysRevLett.103.062001} {\bibfield  {journal} {\bibinfo  {journal}
  {Phys. Rev. Lett.}\ }\textbf {\bibinfo {volume} {103}},\ \bibinfo {pages}
  {062001} (\bibinfo {year} {2009})}\BibitemShut {NoStop}%
\bibitem [{\citenamefont {Lohse}\ \emph {et~al.}(2018)\citenamefont {Lohse},
  \citenamefont {Schweizer}, \citenamefont {Price}, \citenamefont
  {Zilberberg},\ and\ \citenamefont {Bloch}}]{Lohse2018Exploring}%
  \BibitemOpen
  \bibfield  {author} {\bibinfo {author} {\bibfnamefont {M.}~\bibnamefont
  {Lohse}}, \bibinfo {author} {\bibfnamefont {C.}~\bibnamefont {Schweizer}},
  \bibinfo {author} {\bibfnamefont {H.~M.}\ \bibnamefont {Price}}, \bibinfo
  {author} {\bibfnamefont {O.}~\bibnamefont {Zilberberg}}, \ and\ \bibinfo
  {author} {\bibfnamefont {I.}~\bibnamefont {Bloch}},\ }\href {\doibase
  10.1038/nature25000} {\bibfield  {journal} {\bibinfo  {journal} {Nature}\
  }\textbf {\bibinfo {volume} {553}},\ \bibinfo {pages} {55} (\bibinfo {year}
  {2018})}\BibitemShut {NoStop}%
\bibitem [{\citenamefont {Zilberberg}\ \emph {et~al.}(2018)\citenamefont
  {Zilberberg}, \citenamefont {Huang}, \citenamefont {Guglielmon},
  \citenamefont {Wang}, \citenamefont {Chen}, \citenamefont {Kraus},\ and\
  \citenamefont {Rechtsman}}]{Kraus2018Photonic}%
  \BibitemOpen
  \bibfield  {author} {\bibinfo {author} {\bibfnamefont {O.}~\bibnamefont
  {Zilberberg}}, \bibinfo {author} {\bibfnamefont {S.}~\bibnamefont {Huang}},
  \bibinfo {author} {\bibfnamefont {J.}~\bibnamefont {Guglielmon}}, \bibinfo
  {author} {\bibfnamefont {M.}~\bibnamefont {Wang}}, \bibinfo {author}
  {\bibfnamefont {K.~P.}\ \bibnamefont {Chen}}, \bibinfo {author}
  {\bibfnamefont {Y.~E.}\ \bibnamefont {Kraus}}, \ and\ \bibinfo {author}
  {\bibfnamefont {M.~C.}\ \bibnamefont {Rechtsman}},\ }\href {\doibase
  10.1038/nature25011} {\bibfield  {journal} {\bibinfo  {journal} {Nature}\
  }\textbf {\bibinfo {volume} {553}},\ \bibinfo {pages} {59} (\bibinfo {year}
  {2018})}\BibitemShut {NoStop}%
\bibitem [{\citenamefont {Zhang}\ and\ \citenamefont {Hu}(2001)}]{Zhang2001A}%
  \BibitemOpen
  \bibfield  {author} {\bibinfo {author} {\bibfnamefont {S.-C.}\ \bibnamefont
  {Zhang}}\ and\ \bibinfo {author} {\bibfnamefont {J.}~\bibnamefont {Hu}},\
  }\href {\doibase 10.1126/science.294.5543.823} {\bibfield  {journal}
  {\bibinfo  {journal} {Science}\ }\textbf {\bibinfo {volume} {294}},\ \bibinfo
  {pages} {823} (\bibinfo {year} {2001})}\BibitemShut {NoStop}%
\bibitem [{\citenamefont {Kraus}\ \emph {et~al.}(2013)\citenamefont {Kraus},
  \citenamefont {Ringel},\ and\ \citenamefont {Zilberberg}}]{Kraus2013Four}%
  \BibitemOpen
  \bibfield  {author} {\bibinfo {author} {\bibfnamefont {Y.~E.}\ \bibnamefont
  {Kraus}}, \bibinfo {author} {\bibfnamefont {Z.}~\bibnamefont {Ringel}}, \
  and\ \bibinfo {author} {\bibfnamefont {O.}~\bibnamefont {Zilberberg}},\
  }\href {\doibase 10.1103/PhysRevLett.111.226401} {\bibfield  {journal}
  {\bibinfo  {journal} {Phys. Rev. Lett.}\ }\textbf {\bibinfo {volume} {111}},\
  \bibinfo {pages} {226401} (\bibinfo {year} {2013})}\BibitemShut {NoStop}%
\bibitem [{\citenamefont {Price}\ \emph {et~al.}(2015)\citenamefont {Price},
  \citenamefont {Zilberberg}, \citenamefont {Ozawa}, \citenamefont
  {Carusotto},\ and\ \citenamefont {Goldman}}]{Price2015Four}%
  \BibitemOpen
  \bibfield  {author} {\bibinfo {author} {\bibfnamefont {H.~M.}\ \bibnamefont
  {Price}}, \bibinfo {author} {\bibfnamefont {O.}~\bibnamefont {Zilberberg}},
  \bibinfo {author} {\bibfnamefont {T.}~\bibnamefont {Ozawa}}, \bibinfo
  {author} {\bibfnamefont {I.}~\bibnamefont {Carusotto}}, \ and\ \bibinfo
  {author} {\bibfnamefont {N.}~\bibnamefont {Goldman}},\ }\href {\doibase
  10.1103/PhysRevLett.115.195303} {\bibfield  {journal} {\bibinfo  {journal}
  {Phys. Rev. Lett.}\ }\textbf {\bibinfo {volume} {115}},\ \bibinfo {pages}
  {195303} (\bibinfo {year} {2015})}\BibitemShut {NoStop}%
\bibitem [{\citenamefont {Price}(2020)}]{Price2020Four}%
  \BibitemOpen
  \bibfield  {author} {\bibinfo {author} {\bibfnamefont {H.~M.}\ \bibnamefont
  {Price}},\ }\href {\doibase 10.1103/PhysRevB.101.205141} {\bibfield
  {journal} {\bibinfo  {journal} {Phys. Rev. B}\ }\textbf {\bibinfo {volume}
  {101}},\ \bibinfo {pages} {205141} (\bibinfo {year} {2020})}\BibitemShut
  {NoStop}%
\bibitem [{\citenamefont {Terrier}\ and\ \citenamefont
  {Kunst}(2020)}]{Terrier2020Dissipative}%
  \BibitemOpen
  \bibfield  {author} {\bibinfo {author} {\bibfnamefont {F.}~\bibnamefont
  {Terrier}}\ and\ \bibinfo {author} {\bibfnamefont {F.~K.}\ \bibnamefont
  {Kunst}},\ }\href {\doibase 10.1103/PhysRevResearch.2.023364} {\bibfield
  {journal} {\bibinfo  {journal} {Phys. Rev. Research}\ }\textbf {\bibinfo
  {volume} {2}},\ \bibinfo {pages} {023364} (\bibinfo {year}
  {2020})}\BibitemShut {NoStop}%
\bibitem [{\citenamefont {Jensen}\ and\ \citenamefont
  {Koppe}(1971)}]{Jensen1971Quantum}%
  \BibitemOpen
  \bibfield  {author} {\bibinfo {author} {\bibfnamefont {H.}~\bibnamefont
  {Jensen}}\ and\ \bibinfo {author} {\bibfnamefont {H.}~\bibnamefont {Koppe}},\
  }\href {\doibase http://dx.doi.org/10.1016/0003-4916(71)90031-5} {\bibfield
  {journal} {\bibinfo  {journal} {Ann. Phys.}\ }\textbf {\bibinfo {volume}
  {63}},\ \bibinfo {pages} {586 } (\bibinfo {year} {1971})}\BibitemShut
  {NoStop}%
\bibitem [{\citenamefont {da~Costa}(1981)}]{Costa1981Quantum}%
  \BibitemOpen
  \bibfield  {author} {\bibinfo {author} {\bibfnamefont {R.~C.~T.}\
  \bibnamefont {da~Costa}},\ }\href {\doibase 10.1103/PhysRevA.23.1982}
  {\bibfield  {journal} {\bibinfo  {journal} {Phys. Rev. A}\ }\textbf {\bibinfo
  {volume} {23}},\ \bibinfo {pages} {1982} (\bibinfo {year}
  {1981})}\BibitemShut {NoStop}%
\bibitem [{\citenamefont {Wang}\ and\ \citenamefont
  {Zong}(2016)}]{Wang2016Quantum}%
  \BibitemOpen
  \bibfield  {author} {\bibinfo {author} {\bibfnamefont {Y.-L.}\ \bibnamefont
  {Wang}}\ and\ \bibinfo {author} {\bibfnamefont {H.-S.}\ \bibnamefont
  {Zong}},\ }\href {\doibase https://doi.org/10.1016/j.aop.2015.10.019}
  {\bibfield  {journal} {\bibinfo  {journal} {Ann. Phys.}\ }\textbf {\bibinfo
  {volume} {364}},\ \bibinfo {pages} {68 } (\bibinfo {year}
  {2016})}\BibitemShut {NoStop}%
\bibitem [{\citenamefont {Dandoloff}\ and\ \citenamefont
  {Truong}(2004)}]{Dandoloff2004Quantum}%
  \BibitemOpen
  \bibfield  {author} {\bibinfo {author} {\bibfnamefont {R.}~\bibnamefont
  {Dandoloff}}\ and\ \bibinfo {author} {\bibfnamefont {T.~T.}\ \bibnamefont
  {Truong}},\ }\href {\doibase
  http://dx.doi.org/10.1016/j.physleta.2004.03.050} {\bibfield  {journal}
  {\bibinfo  {journal} {Phys. Lett. A}\ }\textbf {\bibinfo {volume} {325}},\
  \bibinfo {pages} {233 } (\bibinfo {year} {2004})}\BibitemShut {NoStop}%
\bibitem [{\citenamefont {Atanasov}\ \emph {et~al.}(2009)\citenamefont
  {Atanasov}, \citenamefont {Dandoloff},\ and\ \citenamefont
  {Saxena}}]{Dandoloff2009Geometry}%
  \BibitemOpen
  \bibfield  {author} {\bibinfo {author} {\bibfnamefont {V.}~\bibnamefont
  {Atanasov}}, \bibinfo {author} {\bibfnamefont {R.}~\bibnamefont {Dandoloff}},
  \ and\ \bibinfo {author} {\bibfnamefont {A.}~\bibnamefont {Saxena}},\ }\href
  {\doibase 10.1103/PhysRevB.79.033404} {\bibfield  {journal} {\bibinfo
  {journal} {Phys. Rev. B}\ }\textbf {\bibinfo {volume} {79}},\ \bibinfo
  {pages} {033404} (\bibinfo {year} {2009})}\BibitemShut {NoStop}%
\bibitem [{\citenamefont {Takagi}\ and\ \citenamefont
  {Tanzawa}(1992)}]{Takagi1992}%
  \BibitemOpen
  \bibfield  {author} {\bibinfo {author} {\bibfnamefont {S.}~\bibnamefont
  {Takagi}}\ and\ \bibinfo {author} {\bibfnamefont {T.}~\bibnamefont
  {Tanzawa}},\ }\href {\doibase 10.1143/ptp/87.3.561} {\bibfield  {journal}
  {\bibinfo  {journal} {Prog. Theo. Phys.}\ }\textbf {\bibinfo {volume} {87}},\
  \bibinfo {pages} {561} (\bibinfo {year} {1992})}\BibitemShut {NoStop}%
\bibitem [{\citenamefont {Gromov}\ \emph {et~al.}(2016)\citenamefont {Gromov},
  \citenamefont {Jensen},\ and\ \citenamefont {Abanov}}]{Gromov2016Boundary}%
  \BibitemOpen
  \bibfield  {author} {\bibinfo {author} {\bibfnamefont {A.}~\bibnamefont
  {Gromov}}, \bibinfo {author} {\bibfnamefont {K.}~\bibnamefont {Jensen}}, \
  and\ \bibinfo {author} {\bibfnamefont {A.~G.}\ \bibnamefont {Abanov}},\
  }\href {\doibase 10.1103/PhysRevLett.116.126802} {\bibfield  {journal}
  {\bibinfo  {journal} {Phys. Rev. Lett.}\ }\textbf {\bibinfo {volume} {116}},\
  \bibinfo {pages} {126802} (\bibinfo {year} {2016})}\BibitemShut {NoStop}%
\bibitem [{\citenamefont {Haldane}()}]{Haldane2009Hall}%
  \BibitemOpen
  \bibfield  {author} {\bibinfo {author} {\bibfnamefont {F.~D.~M.}\
  \bibnamefont {Haldane}},\ }\href {https://arxiv.org/abs/0906.1854} {\bibinfo
  {journal} {arXiv:0906.1854}\ }\BibitemShut {NoStop}%
\bibitem [{\citenamefont {Liu}\ \emph {et~al.}(2017)\citenamefont {Liu},
  \citenamefont {Yang},\ and\ \citenamefont {Li}}]{Liu2017Curvature}%
  \BibitemOpen
\bibfield  {journal} {  }\bibfield  {author} {\bibinfo {author} {\bibfnamefont
  {Q.~H.}\ \bibnamefont {Liu}}, \bibinfo {author} {\bibfnamefont
  {X.}~\bibnamefont {Yang}}, \ and\ \bibinfo {author} {\bibfnamefont
  {Z.}~\bibnamefont {Li}},\ }\href@noop {} {\  (\bibinfo {year} {2017})},\
  \Eprint {http://arxiv.org/abs/1709.07299} {arXiv:1709.07299} \BibitemShut
  {NoStop}%
\bibitem [{\citenamefont {Read}(2009)}]{Read2009NonAbelian}%
  \BibitemOpen
  \bibfield  {author} {\bibinfo {author} {\bibfnamefont {N.}~\bibnamefont
  {Read}},\ }\href {\doibase 10.1103/PhysRevB.79.045308} {\bibfield  {journal}
  {\bibinfo  {journal} {Phys. Rev. B}\ }\textbf {\bibinfo {volume} {79}},\
  \bibinfo {pages} {045308} (\bibinfo {year} {2009})}\BibitemShut {NoStop}%
\bibitem [{\citenamefont {de~Lima}\ and\ \citenamefont
  {Filgueiras}(2012)}]{Lima2012Integer}%
  \BibitemOpen
  \bibfield  {author} {\bibinfo {author} {\bibfnamefont {A.}~\bibnamefont
  {de~Lima}}\ and\ \bibinfo {author} {\bibfnamefont {C.}~\bibnamefont
  {Filgueiras}},\ }\href {\doibase 10.1140/epjb/e2012-30766-9} {\bibfield
  {journal} {\bibinfo  {journal} {Eur. Phys. J. B}\ }\textbf {\bibinfo {volume}
  {85}},\ \bibinfo {pages} {401} (\bibinfo {year} {2012})}\BibitemShut
  {NoStop}%
\bibitem [{\citenamefont {Filgueiras}\ and\ \citenamefont
  {Silva}(2015)}]{Filgueiras20152DEG}%
  \BibitemOpen
  \bibfield  {author} {\bibinfo {author} {\bibfnamefont {C.}~\bibnamefont
  {Filgueiras}}\ and\ \bibinfo {author} {\bibfnamefont {E.~O.}\ \bibnamefont
  {Silva}},\ }\href {\doibase https://doi.org/10.1016/j.physleta.2015.06.035}
  {\bibfield  {journal} {\bibinfo  {journal} {Physics Letters A}\ }\textbf
  {\bibinfo {volume} {379}},\ \bibinfo {pages} {2110 } (\bibinfo {year}
  {2015})}\BibitemShut {NoStop}%
\bibitem [{\citenamefont {de~Lima}\ \emph {et~al.}(2013)\citenamefont
  {de~Lima}, \citenamefont {Poux}, \citenamefont {Assafr{\~a}o},\ and\
  \citenamefont {Filgueiras}}]{Lima2013Screw}%
  \BibitemOpen
  \bibfield  {author} {\bibinfo {author} {\bibfnamefont {A.~G.}\ \bibnamefont
  {de~Lima}}, \bibinfo {author} {\bibfnamefont {A.}~\bibnamefont {Poux}},
  \bibinfo {author} {\bibfnamefont {D.}~\bibnamefont {Assafr{\~a}o}}, \ and\
  \bibinfo {author} {\bibfnamefont {C.}~\bibnamefont {Filgueiras}},\ }\href
  {\doibase 10.1140/epjb/e2013-40160-x} {\bibfield  {journal} {\bibinfo
  {journal} {Eur. Phys. J. B}\ }\textbf {\bibinfo {volume} {86}},\ \bibinfo
  {pages} {485} (\bibinfo {year} {2013})}\BibitemShut {NoStop}%
\bibitem [{\citenamefont {Uchida}\ and\ \citenamefont
  {Tonomura}(2010)}]{Uchida2010Generation}%
  \BibitemOpen
  \bibfield  {author} {\bibinfo {author} {\bibfnamefont {M.}~\bibnamefont
  {Uchida}}\ and\ \bibinfo {author} {\bibfnamefont {A.}~\bibnamefont
  {Tonomura}},\ }\href {\doibase 10.1038/nature08904} {\bibfield  {journal}
  {\bibinfo  {journal} {Nature}\ }\textbf {\bibinfo {volume} {464}},\ \bibinfo
  {pages} {737} (\bibinfo {year} {2010})}\BibitemShut {NoStop}%
\bibitem [{\citenamefont {Verbeeck}\ \emph {et~al.}(2010)\citenamefont
  {Verbeeck}, \citenamefont {Tian},\ and\ \citenamefont
  {Schattschneider}}]{Verbeeck2010Production}%
  \BibitemOpen
  \bibfield  {author} {\bibinfo {author} {\bibfnamefont {J.}~\bibnamefont
  {Verbeeck}}, \bibinfo {author} {\bibfnamefont {H.}~\bibnamefont {Tian}}, \
  and\ \bibinfo {author} {\bibfnamefont {P.}~\bibnamefont {Schattschneider}},\
  }\href {\doibase 10.1038/nature09366} {\bibfield  {journal} {\bibinfo
  {journal} {Nature}\ }\textbf {\bibinfo {volume} {467}},\ \bibinfo {pages}
  {301} (\bibinfo {year} {2010})}\BibitemShut {NoStop}%
\bibitem [{\citenamefont {Silenko}\ \emph {et~al.}(2017)\citenamefont
  {Silenko}, \citenamefont {Zhang},\ and\ \citenamefont
  {Zou}}]{Silenko2017Manipulating}%
  \BibitemOpen
  \bibfield  {author} {\bibinfo {author} {\bibfnamefont {A.~J.}\ \bibnamefont
  {Silenko}}, \bibinfo {author} {\bibfnamefont {P.}~\bibnamefont {Zhang}}, \
  and\ \bibinfo {author} {\bibfnamefont {L.}~\bibnamefont {Zou}},\ }\href
  {\doibase 10.1103/PhysRevLett.119.243903} {\bibfield  {journal} {\bibinfo
  {journal} {Phys. Rev. Lett.}\ }\textbf {\bibinfo {volume} {119}},\ \bibinfo
  {pages} {243903} (\bibinfo {year} {2017})}\BibitemShut {NoStop}%
\bibitem [{\citenamefont {Lloyd}\ \emph {et~al.}(2017)\citenamefont {Lloyd},
  \citenamefont {Babiker}, \citenamefont {Thirunavukkarasu},\ and\
  \citenamefont {Yuan}}]{Lloyd2017Electron}%
  \BibitemOpen
  \bibfield  {author} {\bibinfo {author} {\bibfnamefont {S.~M.}\ \bibnamefont
  {Lloyd}}, \bibinfo {author} {\bibfnamefont {M.}~\bibnamefont {Babiker}},
  \bibinfo {author} {\bibfnamefont {G.}~\bibnamefont {Thirunavukkarasu}}, \
  and\ \bibinfo {author} {\bibfnamefont {J.}~\bibnamefont {Yuan}},\ }\href
  {\doibase 10.1103/RevModPhys.89.035004} {\bibfield  {journal} {\bibinfo
  {journal} {Rev. Mod. Phys.}\ }\textbf {\bibinfo {volume} {89}},\ \bibinfo
  {pages} {035004} (\bibinfo {year} {2017})}\BibitemShut {NoStop}%
\end{thebibliography}%

\end{document}